\documentclass[11pt]{article}
\usepackage[english]{babel}
\usepackage{amssymb,amsmath}

\numberwithin{equation}{section}

\newtheorem{definition}{Definition}[section]
\newtheorem{proposition}{Proposition}[section]
\newtheorem{remark}{Remark}[section]
\newtheorem{lemma}{Lemma}[section]
\newtheorem{theorem}{Theorem}[section]
\newtheorem{corollary}{Corollary}[section]

\newcommand{\tr}{\mathrm{tr}}

\newcommand{\spa}{\hspace{-2mm}}
\newcommand{\tre}{\hspace{0.3mm}}
\newcommand{\quattro}{\hspace{0.4mm}}
\newcommand{\cinque}{\hspace{0.5mm}}
\newcommand{\sei}{\hspace{0.6mm}}
\newcommand{\sette}{\hspace{0.7mm}}
\newcommand{\otto}{\hspace{0.8mm}}
\newcommand{\nove}{\hspace{0.9mm}}
\newcommand{\mtre}{\hspace{-0.3mm}}

\newcommand{\mcinque}{\hspace{-0.5mm}}
\newcommand{\msei}{\hspace{-0.6mm}}
\newcommand{\msette}{\hspace{-0.7mm}}
\newcommand{\motto}{\hspace{-0.8mm}}

\newcommand{\prm}{\mathsf{p}}
\newcommand{\errep}{\mathbb{R}^{\mbox{\tiny $+$}}}
\newcommand{\erreps}{\mathbb{R}_{\hspace{0.3mm}\ast}^{\mbox{\tiny $+$}}}
\newcommand{\erre}{\mathbb{R}}
\newcommand{\erred}{\dot{\mathbb{R}}}
\newcommand{\erredn}{\dot{\mathbb{R}}^n}
\newcommand{\ccc}{\mathbb{C}}
\newcommand{\conv}{\star}
\newcommand{\ot}{\mathrm{o}(t)}
\newcommand{\de}{\mathrm{d}}
\newcommand{\eee}{\mathrm{e}}
\newcommand{\bsp}{\mathfrak{X}}
\newcommand{\csg}{\mathfrak{C}}
\newcommand{\opa}{\hat{A}}
\newcommand{\opb}{\hat{B}}
\newcommand{\ops}{\hat{S}}
\newcommand{\con}{\mathsf{C}}
\newcommand{\cz}{\mathsf{C}_0}
\newcommand{\czc}{\mathsf{C}_{\mathrm{c}}}
\newcommand{\cd}{\mathsf{C}^2}
\newcommand{\cdz}{\mathsf{C}^2_0}
\newcommand{\cdc}{\mathsf{C}^2_{\mathrm{c}}}

\newcommand{\bcinf}{\mathsf{BC}^\infty}

\newcommand{\cinfc}{\mathsf{C}^\infty_{\mathrm{c}}}

\newcommand{\supn}{\|_{\mathrm{sup}}}
\newcommand{\norm}{\|_{\diamond}}
\newcommand{\dom}{\mathrm{Dom}}
\newcommand{\ajk}{a^{jk}}
\newcommand{\bj}{b^j}
\newcommand{\tajk}{\tilde{a}^{jk}}
\newcommand{\tbj}{\tilde{b}^j}
\newcommand{\cj}{c^j(\eta)}
\newcommand{\cjm}{c^j(\eta_m)}

\newcommand{\funct}{\mathsf{F}_{\hspace{-0.3mm}t;\hspace{0.3mm}x}}

\newcommand{\hh}{\mathcal{H}}

\newcommand{\oph}{\hat{H}}
\newcommand{\gene}{\mathfrak{L}}
\newcommand{\geneg}{\mathfrak{G}}

\newcommand{\smis}{\mathcal{M}^1(G)}
\newcommand{\misd}{\mathcal{D}(G)}
\newcommand{\lie}{\mathrm{Lie}\hspace{0.3mm}(G)}
\newcommand{\pu}{\pi_U}
\newcommand{\opx}{\hat{X}}
\newcommand{\opy}{\hat{Y}}

\newcommand{\twi}{\mathfrak{S}_\mu^{U}}
\newcommand{\twihs}{\check{\mathfrak{S}}_\mu^{U}}
\newcommand{\twit}{\mathfrak{S}_{\mu_t}^{U}}
\newcommand{\twiths}{\check{\mathfrak{S}}_{\mu_t}^{U}}
\newcommand{\twis}{\mathfrak{S}_{t}}
\newcommand{\twishs}{\check{\mathfrak{S}}_{t}}

\newcommand{\dwi}{\mathfrak{D}_\mu^{U}}

\newcommand{\bH}{\mathcal{B}(\mathcal{H})}
\newcommand{\brH}{\mathcal{B}_{\mathbb{R}}(\mathcal{H})}
\newcommand{\mut}{\mu_t}
\newcommand{\dime}{\mathtt{N}}
\newcommand{\ima}{\mathrm{i}}

\newcommand{\mat}{\mathsf{M}}

\newcommand{\idm}{I_{\mathtt{M}}}
\newcommand{\idime}{I_{\dime}}
\newcommand{\cm}{\mathbb{C}^{\mathtt{M}}}
\newcommand{\cpm}{\mathfrak{F}}
\newcommand{\ope}{\hat{E}}
\newcommand{\opf}{\hat{F}}
\newcommand{\opk}{\hat{K}}
\newcommand{\opl}{\hat{L}}
\newcommand{\ranglehs}{\rangle_{\mathrm{HS}}}

\newcommand{\Gst}{{G_{\ast}}}
\newcommand{\rest}{{\mathfrak{R}\hspace{0.3mm}}}
\newcommand{\ran}{\mathrm{Ran}\hspace{0.3mm}}
\newcommand{\dimu}{{\mathtt{D}}}
\newcommand{\fast}{^{\phantom{\ast}}}

\newcommand{\sunig}{\mathrm{SU}\hspace{0.3mm}}

\newcommand{\vu}{\mathsf{V}_U}

\newcommand{\prop}{\mathfrak{P}}
\newcommand{\xe}{{\hspace{0.2mm}\bar{x}}}

\newcommand{\sopa}{\mathfrak{A}}
\newcommand{\ah}{\ima\hspace{0.2mm}\brH}

\newcommand{\infgen}{\mathfrak{I}}

\newcommand{\trc}{\mathcal{B}_1(\mathcal{H})}
\newcommand{\den}{\mathcal{N}}

\newcommand{\cpt}{\mathsf{DM}\hspace{0.3mm}(\mathcal{H})}
\newcommand{\cptru}{\mathsf{DM}_{\mathsf{ru}}(\mathcal{H})}
\newcommand{\dmru}{\mathsf{DM}_{\mathsf{ru}}}
\newcommand{\trn}{\|_{\mathrm{tr}}}

\newcommand{\emme}{\mathsf{m}\hspace{0.3mm}}

\newcommand{\wlim}{\mbox{w-}\hspace{-0.7mm}\lim}

\newcommand{\slim}{\mbox{s\hspace{0.3mm}-}\hspace{-0.7mm}\lim}

\newcommand{\hs}{\mathcal{B}_2(\mathcal{H})}
\newcommand{\norhs}{\|_{\mbox{\tiny HS}}^{\phantom{x}}}
\newcommand{\norhssq}{\|_{\mbox{\tiny HS}}}

\newcommand{\compe}{\mathcal{K}_e}

\newcommand{\supp}{\mathrm{supp}\hspace{0.3mm}}

\newcommand{\urep}{{U\hspace{-0.5mm}\vee\hspace{-0.5mm} U}}
\newcommand{\rep}{{\underline{U\hspace{-0.5mm}\vee\hspace{-0.5mm} U\hspace{-0.8mm}}\hspace{0.8mm}}}

\newcommand{\supops}{\mathcal{L}(\mathcal{H})}
\newcommand{\dsupops}{\mathcal{L}^\prime(\mathcal{H})}
\newcommand{\randu}{\mathfrak{U}\hspace{0.3mm}}

\newcommand{\vk}{V_k^{\phantom{\ast}}}
\newcommand{\vkast}{V_k^{\ast}}
\newcommand{\cardi}{\mathrm{card}\hspace{0.3mm}}
\newcommand{\cardin}{\mathsf{c}(\dime)}
\newcommand{\qds}{\mathfrak{Q}_t}

\newcommand{\randue}{\mathfrak{U}_{\hspace{0.5mm}\eta}^{\hspace{0.3mm}U}}
\newcommand{\randuem}{\mathfrak{U}_{\hspace{0.5mm}\eta_m}^{\hspace{0.3mm}U}}

\newcommand{\genera}{\gene\big(U,\{\mut\}\big)}
\newcommand{\generag}{\mathfrak{G}\big(U,\{\mut\}\big)}
\newcommand{\generaw}{\mathfrak{W}\big(U,\{\mut\}\big)}
\newcommand{\generawo}{\mathfrak{W}_0\big(U,\{\mut\}\big)}
\newcommand{\generam}{\gene\big(U,\{\mu_{t;\tre m}\}\big)}

\newcommand{\tk}{{t_k}}

\newcommand{\bumpu}{\beta_{1}}

\newcommand{\bumpm}{\beta_{m}}
\newcommand{\precomp}{\mathcal{K}^{\circ}}
\newcommand{\comp}{\mathcal{K}^{\phantom{\circ}}}
\newcommand{\ocomp}{\mathcal{O}\hspace{-0.5mm}^{\phantom{\circ}}}

\newcommand{\co}{\mathrm{co}\hspace{0.3mm}}
\newcommand{\clco}{\overline{\mathrm{co}}\hspace{0.3mm}}
\newcommand{\cone}{\mathrm{cone}\hspace{0.3mm}}
\newcommand{\clcone}{\overline{\mathrm{cone}}\hspace{0.3mm}}
\newcommand{\zercone}{{\mathrm{cone}}_0}

\newcommand{\clcocone}{\overline{\mbox{\rm co-cone}}\hspace{0.3mm}}

\newcommand{\coco}{\mathcal{C}}
\newcommand{\coconeh}{\mathcal{C}(\mathcal{H})}

\newcommand{\uni}{\mathcal{U}}
\newcommand{\suni}{\mathcal{V}}
\newcommand{\csuni}{\overline{\mathcal{V}}}
\newcommand{\vvv}{{\mathcal{V}\hspace{-0.5mm}\vee\hspace{-0.5mm} \mathcal{V}}}
\newcommand{\cvvv}{{\overline{\mathcal{V}}\hspace{-0.5mm}\vee\hspace{-0.5mm}\overline{\mathcal{V}}}}

\newcommand{\matri}{\mathtt{M}\tre}

\newcommand{\ug}{U(G)}
\newcommand{\cug}{\overline{U(G)}}

\newcommand{\subs}{\mathcal{S}}

\newcommand{\generau}{\mathcal{G}(U)}
\newcommand{\lineal}{\mathcal{G}_0(U)}
\newcommand{\generauno}{\mathcal{G}_1(U)}

\newcommand{\proju}{\hat{P}_U^{\phantom{1}}}
\newcommand{\projinv}{\hat{P}_U^{-1}}

\newcommand{\gumt}{G_0(U,\{\mu_t\})}
\newcommand{\gu}{G_0(U)}
\newcommand{\subg}{\breve{G}}
\newcommand{\subre}{\breve{U}}

\newcommand{\Gext}{G_{\mathrm{ext}}}

\begin{document}

\title{Brownian motion on Lie groups and open quantum systems}

\author{
P. Aniello$^{\ast\hspace{0.5mm}\ddagger}$, A. Kossakowski$^\dagger$,
G. Marmo$^\ast$ and
F. Ventriglia$^\ast$  \vspace{2mm}\\
\small \it $^\ast$ Dipartimento di Scienze Fisiche dell'Universit\`a
di Napoli `Federico
II'\\ \small \it and Istituto Nazionale di Fisica Nucleare (INFN), Sezione di Napoli, \\
\small \it Complesso Universitario di Monte S.\ Angelo, via Cintia,
I-80126 Napoli, Italy \\
\small \it $^\ddagger$ Facolt\`a di Scienze Biotecnologiche,
Universit\`a di Napoli `Federico II', Napoli, Italy
\\ \small \it $^\dagger$ MECENAS, Universit\`a
di Napoli `Federico II', via Mezzocannone 8, I-80134 Napoli, Italy }

\maketitle

\begin{abstract}
\noindent We study the \emph{twirling semigroups} of
(super)operators, namely, certain quantum dynamical semigroups that
are associated, in a natural way, with the pairs formed by a
projective representation of a locally compact group and a
convolution semigroup of probability measures on this group. The
link connecting this class of semigroups of operators with
(classical) Brownian motion is clarified. It turns out that every
twirling semigroup associated with a finite-dimensional
representation is a random unitary semigroup, and, conversely, every
random unitary semigroup arises as a twirling semigroup. Using
standard tools of the theory of convolution semigroups of measures
and of convex analysis, we provide a complete characterization of
the infinitesimal generator of a twirling semigroup associated with
a finite-dimensional unitary representation of a Lie group.
\end{abstract}

%------------------------------------------------------------------------------
\section{Introduction}
\label{intro}
%------------------------------------------------------------------------------

The theory of Brownian motion and its several ramifications form an
evergreen area of research of physics and mathematics. The
interesting history of this subject would deserve a whole article
\emph{per se}; hence, we will content ourselves with recalling just
a few salient facts related to our present contribution. The first
investigations of Brownian motion on a Lie group --- and, more
generally, of probability theory on groups --- seem to be due to
Perrin~{\cite{Perrin}}, who studied Brownian motion on the rotation
group $\mathrm{SO}(3)$, and, later, to L\'evy~{\cite{Levy}} who
provided the first theoretical treatment of probability measures on
$\mathrm{U}(1)$ (also consider the early work of von
Mises~{\cite{Mises}} who, studying the atomic weights, introduced a
normal distribution on the torus). These investigations paved the
way to an extensive study of probability theory on locally compact
groups (started in the 1940s); see the classical
references~{\cite{Grenander,Heyer}}, and the rich bibliography
therein. In particular, fundamental and systematic contributions to
the theory of Brownian motion on Lie groups are due to
Ito~{\cite{Ito}}, Yosida~{\cite{Yosida}} and Hunt~{\cite{Hunt}}.

In~{1966}, Nelson showed that there is a remarkable link connecting
(classical) Brownian motion and the Schr\"odinger
equation~{\cite{Nelson-pap}}. Assuming that a particle of mass $m$
is subject to a Brownian motion with diffusion coefficient
$\hbar/2m$ (and no friction), and using the well known relation
between the particle probability density and the quantum-mechanical
wave function, he was able to derive (formally) the Schr\"odinger
equation.

A different association of the evolution of a quantum system with
Brownian motion was proposed, later on, by
Kossakowski~{\cite{Kossakowski}}. In the pioneering times of the
theory of \emph{open} quantum systems~{\cite{Breuer}} --- a complete
definition of quantum dynamical semigroups and the
Gorini-Kossakowski-Lindblad-Sudarshan classification of the
infinitesimal generators~{\cite{Gorini,Lindblad}} had not been
established yet
--- he observed that there is a class
of semigroups of (super)operators --- acting in a space
of trace class operators --- that are generated, in a natural
way, by the pairs of the type $(U,\{\mut\}_{t\in\errep})$, where $U$ is a
representation of a group $G$ and $\{\mut\}_{t\in\errep}$ is a
convolution semigroup of measures on $G$. In particular, he
considered the case where $G$ is a Lie group and
$\{\mut\}_{t\in\errep}$ is what we call nowadays a \emph{Gaussian}
semigroup of measures (see Sect.~{\ref{groups}}). This class of
convolution semigroups of measures describe the statistical properties of
Brownian motion on $G$ (the natural generalization of the ordinary
Brownian motion).

The aim of the present contribution is to provide a rigorous study
of the above mentioned class of semigroups of superoperators ---
that we will call
\emph{twirling semigroups} --- without restrictions on the
convolutions semigroups of measures considered. In particular, in
the case where $G$ is a Lie group, we will not assume, in general,
to deal with Gaussian semigroups of measures. We will prove that
every twirling semigroup is a \emph{quantum dynamical semigroup}~{\cite{Holevo}}, and, in
the case where $G$ is a Lie group and $U$ is a finite-dimensional
unitary representation, we will provide a complete characterization
of the infinitesimal generators of the twirling semigroups associated with $U$.

Like many other mathematical objects having a `natural' definition,
it turns out that twirling semigroups arise in the study of various
physical contexts. For instance, the analysis of the infinitesimal generators of
the twirling semigroups reveals that this class of semigroups of superoperators
includes, in particular,
the semigroups describing the dynamics of a
finite-dimensional system with a purely random Gaussian stochastic
Hamiltonian~{\cite{Go-Kos}}, and the reduced dynamics of a
finite-dimensional system in the limit of singular coupling to a
reservoir at infinite temperature~{\cite{Fri-Go}}.

The twirling semigroups associated with the defining representation
of the group $\sunig(\dime)$ have been studied by K\"ummerer and
Maassen~{\cite{Kummerer}}, with the aim of characterizing the
dilations of dynamical semigroups that are `essentially
commutative'.

Our interest in twirling semigroups is also motivated by possible
applications in the field of quantum computation and
information~{\cite{Nielsen}}, where, usually, finite-dimensional
quantum systems are considered. In fact, it is well known that a
relevant class of `quantum channels' is formed by the so-called
\emph{random unitary maps}, i.e., by those completely positive
trace-preserving maps that can be expressed as convex superpositions
of unitary transformations. Gregoratti and
Werner~{\cite{Gregoratti}} have given a remarkable characterization
of this class of maps: they are the only quantum channels that enjoy
the property of being \emph{perfectly corrigible} by using, as the
only side-resource, classical information obtained form the
environment. Smolin, Verstraete and Winter~{\cite{Smolin}} have
conjectured that asymptotically many copies of any unital quantum
channel (a quantum bistochastic map~{\cite{Bengtsson}}) --- random
unitary maps form a subset of the set of unital channels --- may be
arbitrarily well approximated by a random unitary map. This
conjecture, if proved, would be a `quantum counterpart' of the
Birkhoff-von Neumann theorem~{\cite{Birkhoff}} on bistochastic
matrices. Recently, Mendl and Wolf~{\cite{Mendl}} have studied the
relation between the set of unital channels and the subset of random
unitary maps, and verified the conjecture in special cases. Other
recent investigations of random unitary maps include applications to
quantum cryptography~{\cite{Bouda}} and quantum state
reconstruction~{\cite{Merkel}}.

It is therefore an interesting and natural issue to characterize the
\emph{random unitary semigroups}, i.e., the quantum dynamical
semigroups consisting of random unitary maps. But it turns out that
--- in the case of a finite-dimensional quantum system ---
there is a precise relation between random unitary semigroups and
twirling semigroups: indeed, every twirling semigroup is a random
unitary semigroup --- see Sect.~{\ref{groups}}
--- and, conversely, it can be shown that every random unitary semigroup arises as a
twirling semigroup. Thus, it is likely that our results --- in addition to their
intrinsic theoretical interest --- may find useful applications in the context
of quantum information.

The paper is organized as follows. In Sect.~{\ref{basic}}, for the
reader's convenience, we will recall some mathematical facts that
are fundamental in the rest of the paper, and we will set the main
definitions and notations. Some further notations will be introduced
later on, closer to the place where they are used. Next, in
Sect.~{\ref{errenne}}, we will briefly discuss the group-theoretical
framework underlying the description of the statistical properties
of `standard' Brownian motion. This should help the reader to
achieve a clearer understanding of the general framework. The main
object of our investigation --- the twirling semigroups --- will be
introduced in Sect.~{\ref{twirling}}, were the basic properties of
these semigroups of superoperators will be studied. In
Sect.~{\ref{groups}}, we will focus on the case of twirling
semigroups associated with finite-dimensional representations of Lie
groups. As already mentioned, this case is relevant for applications
to quantum information. Eventually, in Sect.~{\ref{conclusions}}, a
few conclusions will be drawn.

%------------------------------------------------------------------------------
\section{Definitions, basic known facts and notations}
\label{basic}
%------------------------------------------------------------------------------

In this section, we will fix the main notations, and recall some
basic definitions and results that will be useful in the rest of the
paper. We will be rather concise and, for further details, we invite
the reader to consult the standard references {\cite{Folland}}
(functional analysis and basics in probability theory),
\cite{Hille,Yosida-book} (semigroups of operators),
\cite{Varadarajan,Raja} (Lie groups, representation theory),
\cite{Grenander,Heyer} (probability theory on groups).

Let $\bsp$ be a separable real or complex Banach space. Denoting by
$\errep$ the set of non-negative real numbers (the set of strictly
positive real numbers will be denoted by $\erreps$), a family
$\{\csg_t\}_{t\in\errep}$ of bounded linear operators in $\bsp$ is
said to be a (one-parameter) \emph{semigroup of operators} if the
following conditions are satisfied:
\begin{enumerate}
\item $\csg_t\cinque \csg_s = \csg_{t+s}$,
$\forall\quattro t,s\ge0$ \ (one-parameter semigroup property);

\item $\csg_0 = I$;

\item $\lim_{t\downarrow 0}\|\csg_t\cinque\zeta -\zeta\|=0$,
$\forall\quattro\zeta\in\bsp$, i.e., $\slim_{t\downarrow
0}\csg_t=I$ \  (strong right continuity at $t=0$).
\end{enumerate}
Here and throughout the paper, $I$ is the identity operator.
According to a classical result
--- see~{\cite{Hille}}
--- the previous conditions imply that the map $\errep\ni t\mapsto
\csg_t\in\bsp$ is strongly continuous.
Moreover~{\cite{Yosida-book}}, the last condition is equivalent to
the assumption that $\wlim_{t\downarrow 0}\csg_t=I$ (weak limit). A
semigroup of operators $\{\csg_t\}_{t\in\errep}$ is said to be a
\emph{contraction semigroup} if, in addition to the previous
hypotheses, it satisfies
\begin{enumerate}
\setcounter{enumi}{3}
\item $\|\csg_t\|\le 1$, $\forall\quattro t >0$.
\end{enumerate}

A semigroup of operators $\{\csg_t\}_{t\in\errep}$ admits a densely
defined \emph{infinitesimal generator}, namely, the closed linear
operator $\sopa$ in $\bsp$ defined by
\begin{equation}
\dom(\sopa):=\Big\{\zeta\in\bsp\colon \exists\cinque
\lim_{t\downarrow 0}
t^{-1}\big(\csg_t\cinque\zeta-\zeta\big)\Big\},\ \ \
\sopa\cinque\zeta := \lim_{t\downarrow 0}
t^{-1}\big(\csg_t\cinque\zeta-\zeta\big),\
\forall\quattro\zeta\in\dom(\sopa).
\end{equation}

Let $X$ be a locally compact, second countable, Hausdorff
topological space (in short, l.c.s.c.\ space). If $X$ is noncompact,
the symbol $\dot{X}$ will indicate the one-point compactification of
$X$. We will denote by $\cz(X)$ the Banach space of all continuous
$\erre$-valued functions on $X$ vanishing at infinity (hence,
bounded), endowed with the `sup-norm':
\begin{equation}
\|f\supn := \sup_{x\in X} |f(x)|,\ \ \ f\in \cz(X).
\end{equation}
As is well known, $\cz(X)$ is the closure, with respect to the
sup-norm, of the vector space $\czc(X)$ of all continuous
$\erre$-valued functions on $X$ with compact support. If $X$ is
noncompact, the vector space $\cz(X)$ can be immersed in a natural
way in $\con(\dot{X})$, the Banach space of continuous real-valued
functions on $\dot{X}$ (endowed with the sup-norm) --- i.e., setting
$f(\infty)=0$, for all $f\in\cz (X)$ ---  and every function in
$\con(\dot{X})$  can be expressed as the sum of a function in
$\cz(X)$ and a constant function.

We will call a contraction semigroup $\{\csg_t\}_{t\in\errep}$ in the Banach space
$\cz(X)$ a \emph{Markovian semigroup} if it satisfies the conditions
\begin{equation} \label{condi1}
\cz(X)\ni f\ge 0\ \ \ \Rightarrow\ \ \ \csg_t\cinque f\ge 0,\ \forall\quattro t> 0,
\end{equation}
(hence: $\csg_t\cinque f_1\ge\csg_t\cinque f_2$, for $f_1\ge f_2$)
--- thus, for each $x\in X$, the map $\funct\colon \cz(X)\ni
f\mapsto\big(\csg_t\cinque f\big)(x)$ must be a (bounded) positive
functional, with $\|\funct\|\le\|\csg_t\|\le 1$
--- \emph{and}
\begin{equation} \label{condi2}
\sup_{f\in\cz(X),\ 0\le f \le 1} \hspace{-1.5mm}\big(\csg_t\cinque
f\big)(x)=1,\ \ \ \forall\quattro x\in X,\ \forall\quattro t> 0;
\end{equation}
i.e., $\|\funct\|\ge 1$, hence: $\|\funct\|= 1$. Clearly,
condition~{(\ref{condi2})} implies that the contraction semigroup
$\{\csg_t\}_{t\in\errep}$ is such that $\|\csg_t\|=1$, for all $t\ge
0$. Moreover, by the Riesz representation theorem there exists a
unique family $\{\prm_{t;\tre x}\colon t\in\errep,\otto x\in X\}$ of
(regular)\footnote{Recall that in a l.c.s.c.\ space every finite
Borel measure is regular.} probability measures on $X$ such that
\begin{equation} \label{family}
\big(\csg_t\cinque f\big)(x)= \funct \cinque f = \int_X f(y)\;
\de\prm_{t;\tre x}(y), \ \ \ \forall\quattro f\in\cz(X),\
\forall\quattro x\in X,\ \forall\quattro t\ge 0.
\end{equation}

Assume, in particular, that the topological space $X$ is compact.
Then, $1\in\cz(X)$ ($=\con(X)$), and $1=\|\funct\|=\funct\cinque 1$,
for all $x\in X$ and $t\ge 0$. Therefore, in this case,
condition~{(\ref{condi2})} can be replaced by the following:
\begin{equation}
\csg_t\cinque 1 = 1,\ \ \ \forall\quattro t> 0.
\end{equation}

We will denote by $\cdc(\erre^n)$ the vector space of all
$\erre$-valued functions on $\erre^n$, `of class $\mathrm{C}^2$',
with compact support. The completion of this vector space with
respect to the norm
\begin{equation} \label{norma}
\|f\norm := \|f\supn + \sum_{j=1}^{n}
\left\|\frac{\partial}{\partial x^j}\tre f \right\supn +
\sum_{j,k=1}^{n} \left\|\frac{\partial^2}{\partial x^j x^k}\tre f
\right\supn .
\end{equation}
is a real Banach space which will be denoted by $\cdz(\erre^n)$ (it
is clear that $\cdz(\erre^n)\subset\cz(\erre^n)$). Moreover, we will
denote by the symbol $\cd(\erredn)$ the completion with respect to
the norm~{(\ref{norma})} of the real vector space consisting of
linear superpositions of functions in the vector space
$\cdc(\erre^n)$ and the constant functions on $\erre^n$.

Complexifications of some of the real vector spaces of functions
introduced above will also be considered. For instance, we will
consider the complexification $\czc(X;\ccc)$  of the real vector
space $\czc(X)\equiv\czc(X;\erre)$. The notations adopted will be
consistent with this example.

Let $G$ be a locally compact, second countable, Hausdorff
topological group (in short, l.c.s.c.\ group). The symbol $e$ will
denote the identity in $G$, and $\Gst$ the set
$G\smallsetminus\{e\}$.

We will mean by the term {\it projective representation} of $G$ a
Borel projective representation of $G$ in a separable complex
Hilbert space $\mathcal{H}$ (see, for instance, ref.~\cite{Raja},
chapter~{VII}), namely a map $U$ of $G$ into
$\mathcal{U}(\mathcal{H})$  --- the unitary group of $\hh$ --- such
that
\begin{itemize}
\item
$U$ is a weakly Borel map, i.e.\ $G\ni g\mapsto
\langle\phi,U(g)\,\psi\rangle\in\mathbb{C}$ is a Borel function, for
any pair of vectors $\phi,\psi\in\mathcal{H}$;
\item
$U(e)=I$;
\item
denoting by $\mathbb{T}$ the circle group, namely the group of
complex numbers of modulus one, there exists a Borel function $\emme
\colon G\times G\rightarrow\mathbb{T}$ such that
\begin{equation}
U(gh)=\emme (g,h)\,U(g)\,U(h),\ \ \ \forall\hspace{0.5mm} g,h\in G .
\end{equation}
\end{itemize}
The function $\emme$ is called the {\it multiplier associated with}
$U$ (multipliers, however, will play no relevant role in our later
discussion). Clearly, in the case where $\emme\equiv 1$, $U$ is a
standard unitary representation; in this case, according to a well
known result, the hypothesis that the map $U$ is weakly Borel
implies that it is, actually, strongly continuous.

We will denote by $\smis$ the semigroup --- with respect to
convolution of measures\footnote{Recall that for $\mu,\nu\in\smis$
the convolution of $\mu$ with $\nu$ is the measure
$\mu\conv\nu\in\smis$ determined by $\int_G \de \mu\conv\nu (g)\;
f(g) =\int_G  \de \mu(g)\int_G\de \nu (h)\;f(gh)$, for all
$f\in\czc(G;\erre)$.}
--- of all (regular) probability measures on $G$, \emph{endowed with
the weak topology} (which, in $\smis$, coincides with the vague
topology). The symbol $\delta\equiv\delta_e$ will denote the Dirac
measure at $e$, measure that is, of course, the identity in the
semigroup $\smis$. By a \emph{continuous convolution semigroup of
measures} on $G$ we mean a subset $\{\mut\}_{t\in\errep}$ of $\smis$
such that the map $\errep\ni t\mapsto\mut\in\smis$ is a homomorphism
of semigroups and
\begin{equation}
\lim_{t\downarrow 0} \mut = \delta.
\end{equation}
It is a well known fact that this condition implies that the
homomorphism $t\mapsto\mut$ is continuous. Let $\mu$ be a
probability measure in $\smis$. The \emph{probability operator}
associated with $\mu$ is a bounded linear operator $\prop_\mu\colon
\cz(G) \rightarrow\cz(G)$ defined by
\begin{equation}
\big(\prop_\mu\tre f\big)(g) := \int_G f(gh)\; \de\mu(h) = \int_G
f(h)\; \de\mu_g(h),\ \ \ \forall\quattro f\in \cz(G),
\end{equation}
where the probability measure $\mu_g$ is the \emph{$g$-translate} of
the measure $\mu$. Furthermore, the probability operator $\prop_\mu$
is \emph{left-invariant}, i.e.
\begin{equation}
\prop_\mu\sette \ell_g = \ell_g\cinque \prop_\mu,\ \ \
\forall\quattro g\in G
\end{equation}
--- where $\ell_g\colon \cz(G) \rightarrow\cz(G)$ is the isometry
defined by $\ell_g \tre f := f(g(\cdot))$ --- and $\|\prop_\mu\|=1$
(by one of the assertions of the Riesz representation theorem).

A convolution semigroup of measures on $G$ generates, in a natural
way, a contraction semigroup. Precisely, let $\{\mut\}_{t\in\errep}$
be a continuous convolution semigroup of measures on $G$. Then,
setting
\begin{equation} \label{sgasswithm}
\prop_t :=\prop_{\mut},\ t\ge 0,\ \ \ (\prop_0 =I),
\end{equation}
we get a contraction semigroup $\{\prop_t\}_{t\in\errep}$ ---
precisely, a Markovian semigroup --- in the Banach space $\cz(G)$,
which is left-invariant: $\prop_t\sette \ell_g = \ell_g\cinque
\prop_t$, for all $g\in G$ and $t\in\errep$. A semigroup of
operators of the type~{(\ref{sgasswithm})} will be called a
\emph{probability semigroup} on G. Actually, it turns out that
definition~{(\ref{sgasswithm})} establishes a one-to-one
correspondence between the left-invariant Markovian semigroups in
$\cz(G)$ and the continuous convolution semigroups of measures on
$G$ (or the associated probability semigroups).

Let now $G$ be, in particular, a Lie group of dimension $n\ge 1$. We
will denote by $\bcinf(G)$, $\cinfc(G)$ the vector spaces of all
bounded smooth real-valued functions on $G$ and of all smooth
real-valued functions on $G$ with compact support, respectively. For
every basis $\{\xi_1,\ldots,\xi_n\}$ in the Lie algebra $\lie$
(realized as the space of left-invariant vector fields) of $G$,
there exists a relatively compact neighborhood $\mathcal{E}_e$ of
the identity in $G$ and a local chart
\begin{equation}
\left(\mathcal{E}_e;\tre \mathcal{E}_e\ni g\mapsto x^1(g), \ldots,
\mathcal{E}_e\ni g\mapsto x^n(g)\right)
\end{equation}
such that $\exp_G(\sum_{k=1}^n x^k(g)\sei\xi_k)=g$, for all
$g\in\mathcal{E}_e$. Such a local chart is called \emph{a system of
canonical coordinates} (of the first kind) associated with the basis
$\{\xi_1,\ldots,\xi_n\}$. The local maps $g\mapsto x^1(g), \ldots,
g\mapsto x^n(g)$ defined in $\mathcal{E}_e$ can be extended to
suitable real functions
\begin{equation} \label{adapted}
G\ni g\mapsto \xe^1(g)\in\mathbb{R}, \ldots,
G\ni g\mapsto \xe^n(g)\in\mathbb{R},
\end{equation}
belonging to $\cinfc(G)$. We will call such a set of real functions
a \emph{system of adapted coordinates} (based at the identity $e$)
for the Lie group $G$.

Let $U$ be a \emph{smooth unitary} representation\footnote{As is
well known, a continuous homomorphism between Lie groups is
necessarily smooth. Therefore, it would be enough to assume
continuity in order to ensure smoothness.} of the Lie group $G$ in a
\emph{finite-dimensional} Hilbert space $\mathcal{H}$. Then, there
is a unique representation $\pu$ of the Lie algebra $\lie$ in
$\mathcal{H}$ determined by
\begin{equation}
U(\exp_G(\xi))= \eee^{\pu(\xi)}, \ \ \ \forall\quattro \xi\in\lie.
\end{equation}
It is clear that $\pu(\lie)\subset\ah$, with $\ah$ denoting the
finite-dimensional real vector space consisting of all skewadjoint
operators in $\hh$ (accordingly, the real vector space of
selfadjoint operators in $\hh$ will be denoted by $\brH$). We will
adopt the following notation:
\begin{equation} \label{opxs}
\opx_1\equiv \pu(\xi_1),\ldots , \opx_n\equiv \pu(\xi_n).
\end{equation}
Observe that the map $G\ni g\mapsto \eee^{\pu(\xe^1(g) \sei \xi_1 +
\cdots + \xe^n(g) \sei \xi_n)}\in\bH$ is a smooth function such that
\begin{equation}
U(g) =  \eee^{\pu(\xe^1(g) \sei \xi_1 + \cdots + \xe^n(g) \sei
\xi_n)} = \eee^{\xe^1(g) \cinque \opx_1 + \cdots + \xe^n(g) \cinque
\opx_n},\ \ \ \forall\quattro g\in\mathcal{E}_e.
\end{equation}

We will now recall a classical result about left-invariant Markovian
semigroups (probability semigroups)~{\cite{Grenander,Heyer}}. Let
$\Xi\equiv\{\xi_1,\ldots,\xi_n\}$ be a basis in $\lie$. A \emph{Hunt
function} associated with $\Xi$ is real-valued function on $G$ that
verifies the following conditions: it is a function $\Phi$ contained
in $\bcinf(G)$, with $0<\Phi\le 1$, such that
\begin{equation}
\Phi (g)= \sum_{j=1}^n x^j(g)^2,\ \ \ \forall\quattro
g\in\mathcal{E}_e, \ \ \mbox{and}\ \  \Phi(g)=1,\ \forall\quattro
g\in\complement\tre\compe,
\end{equation}
where $\mathcal{E}_e$ is a relatively compact neighborhood of $e$,
$\left(\mathcal{E}_e;\tre \mathcal{E}_e\ni g\mapsto x^1(g), \ldots,
\mathcal{E}_e\ni g\mapsto x^n(g)\right)$ a system of canonical
coordinates (extendable to adapted coordinates denoted as
in~{(\ref{adapted})}) associated with the basis $\Xi$ and $\compe$ a
compact neighborhood of the identity. A \emph{L\'evy measure} $\eta$
is a Radon measure on $\Gst$ satisfying
\begin{equation} \label{levicond}
\int_\Gst \motto \Phi(g)\; \de \eta (g)<\infty,
\end{equation}
for any Hunt function $\Phi$.  Let us denote by $\infgen$ the
infinitesimal generator of a probability semigroup
$\{\prop_t\}_{t\in\errep}$ in $\cz(G)$. Then, the domain of the
operator $\infgen$ contains the vector space $\cdc(G)$, and there
exist real numbers $b^1,\ldots ,b^n$, a positive,\footnote{In the
following, by \emph{positive} we will always mean positive
semidefinite.} symmetric real matrix $\big[\ajk\big]_{j,k=1}^n$ and
a L\'evy measure $\eta$ on $\Gst$ such that
\begin{equation} \label{LKF}
\big(\infgen\tre f\big)(g) = \sum_{j=1}^n \bj\cinque\big(\xi_j\tre
f\big)(g) + \sum_{j,k=1}^n \ajk\cinque\big(\xi_j\tre\xi_k\tre f\big)
(g) + \big(\rest f\big) (g),
\end{equation}
for all $f\in\cdc(G)$, where:
\begin{equation} \label{interm}
\big(\rest f\big) (g) = \int_\Gst \hspace{-1.3mm}\Big( f(gh)- f(g) -
\sum_{j=1}^n \big(\xi_j\tre f\big)(g)\cinque \xe^j(h)\Big) \quattro
\de \eta(h).
\end{equation}
This result is the celebrated \emph{L\'evy-Kintchine formula}. If
$\{\mut\}_{t\in\errep}$ is the continuous convolution semigroup of
measures that generates the probability semigroup
$\{\prop_t\}_{t\in\errep}$, then the L\'evy measure $\eta$ is
uniquely determined by the condition
\begin{equation} \label{caralevi}
\int_\Gst \motto f(g)\; \de
\eta(g)= \lim_{t\downarrow 0}\sette t^{-1}\mtre \int_G f(g)\;\de\mut (g),\ \ \
\forall\quattro f\in \czc (\Gst),\ \ \ (f(e)\equiv 0).
\end{equation}

Conversely --- \emph{given} real numbers $b^1,\ldots ,b^n$, a
positive, symmetric real matrix $\big[\ajk\big]_{j,k=1}^n$ and a
L\'evy measure $\eta$ on $\Gst$ --- one can prove that there is a
contraction semigroup $\{\prop_t\}_{t\in\errep}$ whose infinitesimal
generator satisfies the L\'evy-Kintchine formula~{(\ref{LKF})}.
Therefore, it is natural to call a set $\big\{b^1,\ldots ,b^n;
\big[\ajk\big]_{j,k=1}^n;\eta\big\}$ of the type just described a
\emph{representation kit} (this term is non-standard) of the
contraction semigroup $\{\prop_t\}_{t\in\errep}$.

\begin{remark}
{\rm Since the L\'evy-Kintchine formula~{(\ref{LKF})} has been
written for functions in $\cdc(G)$ --- that is perfectly fit for our
purposes --- we can use the standard Lie derivatives
$\xi_1,\ldots,\xi_n$ of functions on $G$ instead of the `uniform
derivatives' (i.e., derivatives converging in the sup-norm, defined
on suitable Banach spaces), as it is usually done in more general
contexts~{\cite{Grenander,Heyer}}.

A probability semigroup $\{\prop_t\}_{t\in\errep}$ acting in
$\cz(G)\equiv \cz(G;\erre)$ can be extended to $\cz(G;\ccc)$ `by
complexification' and the infinitesimal generator of this extended
semigroup is the complexification of the generator $\infgen$ of
$\{\prop_t\}_{t\in\errep}$. With a slight abuse, we will still
denote by $\infgen$ the complexified generator, and the
L\'evy-Kintchine formula~{(\ref{LKF})} will be understood to hold,
in general, in $\czc(G;\ccc)$.

It is convenient to classify convolution semigroups of measures on Lie groups
according to the behavior of the associated L\'evy measures. We will say that
$\{\mut\}_{t\in\errep}$ is of \emph{regular type} if the associated L\'evy measure
$\eta$ satisfies
\begin{equation} \label{regtype}
\int_\Gst  \sum_{j=1}^n |\xe^j(g)| \; \de \eta (g)<\infty.
\end{equation}
This condition does not depend on the choice of the adapted
coordinates. Note that, if~{(\ref{regtype})} is verified, we have:
\begin{equation} \label{interm-bis}
\big(\rest f\big) (g) = \int_\Gst \hspace{-1.3mm}\Big( f(gh)- f(g)\Big)\quattro \de \eta(h) -
\sum_{j=1}^n \big(\xi_j\tre f\big)(g) \int_\Gst \motto  \xe^j(h)\; \quattro
\de \eta(h).
\end{equation}
We will, moreover, single out a special class of convolution
semigroups of measures of regular type. We will say that the
convolution semigroup of measures $\{\mut\}_{t\in\errep}$ is of the
\emph{first kind} if the associated L\'evy measure $\eta$ on $\Gst$
is finite (hence, satisfies~{(\ref{regtype})}). Otherwise, we will
say that it is a convolution semigroup of measures of the
\emph{second kind}. Clearly, the convolution semigroups of measures
of the second kind that are of nonregular type are characterized by
L\'evy measures satisfying~{(\ref{levicond})} but not the more
stringent condition~{(\ref{regtype})}. ~{$\blacksquare$} }
\end{remark}

Let $\mathcal{A}$ be a $\mathrm{C}^\ast$-algebra. We recall that a
bounded linear map $\Phi\colon \mathcal{A}\rightarrow\mathcal{A}$ is
said to be \emph{completely positive} if the map
$\Phi\otimes\idm\colon\hh\otimes\cm\rightarrow\hh\otimes\cm$ ---
with $\idm$ denoting the identity operator in $\cm$ --- is positive
for any $\mathtt{M}\in\mathbb{N}$. As is well known, in the case
where $\mathcal{A}=\bH$ --- the $\mathrm{C}^\ast$-algebra of all
bounded linear maps in a separable complex Hilbert space $\hh$ ---
and $\dim(\hh)= \dime <\infty$, $\Phi$ is completely positive if and
only if it is $\dime$-positive, i.e.\ $\Phi\otimes\idime$ is
positive. It is also known (see, e.g., ref.~{\cite{Paulsen}}) that
the map $\Phi$ is $\dime$-positive if and only if, for every
$\dime$-tuple $\{\psi_1,\ldots ,\psi_\dime\}$ in $\hh$ and every
$\dime$-tuple
$\big\{\opa_1^{\phantom{\ast}},\ldots,\opa_\dime^{\phantom{\ast}}\big\}$
in $\bH$,
\begin{equation} \label{iff}
\sum_{j,k=1}^{\dime}\big\langle\psi_j,
\Phi\big(\opa_j^\ast\tre\opa_k^{\phantom{\ast}}\big)\psi_k\big\rangle\ge 0.
\end{equation}

%------------------------------------------------------------------------------
\section{The Brownian motion on $\mathbb{R}^n$}
\label{errenne}
%------------------------------------------------------------------------------

The aim of this section is to recall that the statistical properties
of `standard' Brownian motion --- i.e., the Brownian motion on the
Euclidean space $\erre^n$ --- can be expressed, in a natural way, in
the language of one-parameter semigroups of probability measures
(technically, the distributions associated with the Wiener processes
that are the mathematical formalization of Brownian
motion~{\cite{Karatzas}}) and of the associated Markovian
semigroups. In this case ($G=\erre^n$), it will be instructive to
consider a slightly more general mathematical context with respect
to the one considered in Sect.~{\ref{basic}} for introducing the
L\'evy-Kintchine formula~{(\ref{LKF})}. This will help the reader,
in particular, to appreciate the role of the invariance with respect
to translations in our discussion. We will essentially follow the
approach of Nelson's classical book~{\cite{Nelson}}.

As is well known --- see~{\cite{Pathria}} ---
the evolution of the probability distribution of
the position of a Brownian particle (in $\erre^n$, $n\ge 1$),
suspended in a viscous, infinitely extended fluid, can be
regarded as the diffusion through the fluid of a
unit mass initially concentrated in a point, let's say the origin
of $\erre^n$. If the relevant properties of the fluid are assumed to be
invariant with respect to translations and the external forces
acting on the Brownian particle are constant (with respect to space
and time) --- a constant force field that causes a constant
(average) drift velocity of a particle in the fluid~{\cite{Ghez}}
--- then by translating in $\erre^n$ any solution of the equations
governing the diffusion process one must obtain another solution.

Let us formalize mathematically the diffusion process just
described. We will start considering the simplest case: a single
degree of freedom and no drift. Let us consider, then, a family of
probability measures $\{\mut\}_{t\in\erreps}$ on $\erre$ such that
\begin{equation}
\mut \conv \mu_s = \mu_{t+s},\ \ \ t,s\in\erreps,
\end{equation}
where we recall that $\mut \conv \mu_s$ is the convolution of the measure $\mut$ with the neasure
$\mu_s$.
Suppose that, for all $\epsilon >0 $,
\begin{equation} \label{supponi}
\mut (\{y\colon |y|\ge\epsilon\})= \ot,\ \ \ t\downarrow 0.
\end{equation}
Note that this assumption implies, in particular, that
\begin{equation} \label{weaklim}
\lim_{t\downarrow 0} \mut = \delta\ \ \ \mbox{(weakly)}.
\end{equation}
Hence --- setting $\mu_0=\delta$ --- $\{\mut\}_{t\in\errep}$ is a continuous
convolution semigroup of measures on $\erre$. Suppose, moreover,
that the measure $\mut$ is invariant with respect to the
transformation $x\mapsto -x$ ($\Leftrightarrow$ no drift). Then, it
follows that either $\mut =\delta$, for all $t\in\errep$
--- there is no diffusion ---
or, for $t>0$, $\mut$ is absolutely continuous with respect to the
Lebesgue measure on $\erre$ and
\begin{equation}
\de\mut(y)= \wp_t(y)\sei\de y = \frac{1}{\sqrt{4\pi Dt}}\nove
\eee^{-(y^2/4Dt)}\sei\de y,\ \ \ t>0,
\end{equation}
for some $D>0$ (diffusion constant). Thus, the Radon-Nikodym
derivative $\wp_t$ of the measure $\mut$ with respect to the
Lebesgue measure satisfies the diffusion equation
\begin{equation} \label{diffusion}
\frac{\partial}{\partial t}\sette \wp_t(y) = D
\frac{\partial^2}{\partial y^2}\sette \wp_t(y),\ \ \ t>0;
\end{equation}
precisely, it is the fundamental solution of this equation. The
translation-invariant semigroup (probability semigroup)
$\{\prop_t\}_{t\in\errep}$ associated with the semigroup of
probability measures $\{\mut\}_{t\in\errep}$ is given by
\begin{equation}
\big(\prop_t\tre f\big)(x) :=\int_{\erre} f(x+y)\;
\de\mut(y)=\int_{\erre} f(y)\sette\wp_t(y-x)\; \de y, \ \ f\in
\cz(\erre), \ \ t>0,\ \ \ (\prop_0 =I).
\end{equation}
Clearly, for $f\ge 0$ and $t >0$, $\prop_t\tre f$ can be interpreted
as the (expected) concentration, at the time $t$, of a suspension of
Brownian particles with initial ($t=0$) concentration $f$. Note that
one can extend, in a natural way, the domain of the operators in the
semigroup $\{\prop_t\}_{t\in\errep}$ to include linear
superpositions with the constant functions in such a way to obtain a
Markovian semigroup in the Banach space $\con(\erredn)$
($\erredn=\erre^n\cup\infty$). Obviously, this Markovian semigroup
commutes with translations.

Keeping in mind the `elementary case' briefly sketched above, let us
now consider a more general setting. We will focus on the
implications of an assumption of the type~{(\ref{supponi})}, without
assuming, at first, invariance with respect to translations. Then,
let $\{\csg_t\}_{t\in\errep}$ be a Markovian semigroup in the Banach
space $\con(\erredn)$, and let $\sopa$ be the associated
infinitesimal generator. Suppose that
\begin{equation} \label{condi3}
\dom(\sopa)\supset \cdc(\erre^n)
\end{equation}
(a technical condition), and, for all $x\in \erre^n$ and all
$\epsilon >0 $,
\begin{equation} \label{contra}
\prm_{t;\tre x}\big(\{y\in\erre^n\colon |y-x|\ge\epsilon\}\big)=
\ot,\ \ \ t\downarrow 0,
\end{equation}
where $\{\prm_{t;\tre x}\colon t\in\errep,\otto x\in \erredn\}$ is
the family of probability measures determined by~{(\ref{family})},
with $X=\erredn$. Then, one can prove that there are continuous
real-valued functions $\ajk$ and $\bj$ on $\erre^n$,
$j,k=1,\ldots,n$, such that
\begin{equation} \label{stform}
\big(\sopa\tre f\big)(x)= \sum_{j=1}^{n} \bj(x)\cinque
\frac{\partial}{\partial x^j}\tre f(x)  + \sum_{j,k=1}^{n}
\ajk(x)\cinque \frac{\partial^2}{\partial x^j \partial x^k}\tre
f(x),\ \ \ \forall\quattro f\in \cdc(\erre^n),\ \ \ \forall\quattro
x\in \erre^n .
\end{equation}
Moreover, for each $x\in\erre^n$, the matrix
$\left[\ajk(x)\right]_{\hspace{-0.5mm}j,k=1}^n$ is positive, i.e.
\begin{equation}
\sum_{j,k=1}^{n}  \ajk(x)\cinque z_j^{\ast} z_k^{\phantom{\ast}}\ge
0,\ \ \ \forall \quattro z_1,\ldots,z_n\in\ccc.
\end{equation}
As the matrix $\left[\ajk(x)\right]_{\hspace{-0.5mm}j,k=1}^n$ may be
singular, the operator $\sopa$ is not necessarily elliptic. It is
clear that, in the case where the Markovian semigroup
$\{\csg_t\}_{t\in\errep}$ commutes with translations, i.e.
\begin{equation}
\big(\csg_t \tre f\big)(x+(\cdot)) =\csg_t\big(f(x+(\cdot))\big),\ \ \
f\in\con(\erredn),\ \ \ x\in\erre^n,\ \ \ (x+\infty\equiv\infty),
\end{equation}
we have that, for every $x\in\erre^n$,
\begin{equation}
\big(\csg_t\tre f\big)(x) = \int_{\erredn} f(y)\;
\de\prm_{t;\tre x}(y)=\int_{\erredn} f(x+y)\;
\de\prm_{t}(y),\ \ \ \prm_{t}\equiv \prm_{t;\tre 0}.
\end{equation}
Hence, the probability measure $\prm_{t;\tre x}$ is the
$x$-translate of $\prm_{t}$. It is also clear that, in this case, in
formula~{(\ref{stform})} the functions $\ajk$ and $\bj$,
$j,k=1,\ldots,n$, must be constant.

Let now $\{\csg_t\}_{t\in\errep}$ be a Markovian semigroup in the
Banach space $\con(\erredn)$ \emph{that commutes with translations}.
It can be shown that the infinitesimal generator $\sopa$ of such a
semigroup verifies
\begin{equation}
\dom(\sopa)\supset \cd(\erredn).
\end{equation}
Therefore, in this case, condition~{(\ref{condi3})} is automatically
satisfied. If, in addition, for all $\epsilon >0 $, $ \prm_t
(\{y\colon |y|\ge\epsilon\})= \ot$ ($\prm_{t}\equiv \prm_{t;\tre
0}$), for $t\downarrow 0$, then condition~{(\ref{contra})} is
satisfied too (as $\prm_{t;\tre x}$ is the $x$-translate of
$\prm_{t}$), and equation~{(\ref{stform})} holds, in this case with
the real-valued functions $\ajk$ and $\bj$, $j,k=1,\ldots,n$, that
are actually constant (and the matrix
$\left[\ajk\right]_{\hspace{-0.5mm}j,k=1}^n$ positive). We stress
that, in the present paper, we are interested in the case where
$\prm_t(\infty)=0$, for all $t>0$ (`no masses escaping to
infinity').

Let $\{\prop_t\}_{t\in\errep}$ be a translation-invariant Markovian
semigroup in $\cz(\erre^n)$, and let $\{\mut\}_{t\in\errep}$ be the
continuous convolution semigroup of measures that generates this
semigroup. Then, extending the measure $\mut$ to a probability
measure $\prm_t$ on $\erredn$ ($\prm_t(\infty)=0$), one can define a
Markovian semigroup $\{\csg_t\}_{t\in\errep}$ in $\con(\erredn)$
that commutes with translations:
\begin{equation}
\big(\csg_t\tre f\big)(x) :=\int_{\erredn} f(x+y)\; \de\prm_{t}(y),\
\ \ f\in \con(\erredn).
\end{equation}
Assume, moreover, that $\{\mut\}_{t\in\errep}$
satisfies~{(\ref{supponi})}, so that condition~{(\ref{contra})} is
satisfied for the semigroup $\{\csg_t\}_{t\in\errep}$ (as well as
condition~{(\ref{condi3})}). Being $\cz(\erre^n)$ an invariant
subspace for the Markovian semigroup $\{\csg_t\}_{t\in\errep}$, we
can define the linear operator $\infgen\colon
\cz(\erre^n)\cap\dom(\sopa)\ni f\mapsto\sopa f\in\cz(\erre^n)$,
which is precisely the infinitesimal generator of
$\{\prop_t\}_{t\in\errep}$. Thus, from our previous discussion it
follows that
\begin{equation} \label{stform-bis}
\big(\infgen\tre f\big)(x)= \sum_{j=1}^{n} \bj\cinque
\frac{\partial}{\partial x^j}\tre f(x)  + \sum_{j,k=1}^{n}
\ajk\cinque \frac{\partial^2}{\partial x^j \partial x^k}\tre f(x),\
\ \ \forall\quattro f\in \cdc(\erre^n),
\end{equation}
for some real constants $b_1,\ldots,b_n$ and a positive matrix
$\left[\ajk\right]_{\hspace{-0.5mm}j,k=1}^n$. It can be shown,
moreover, that $\infgen$ is uniquely determined
by~{(\ref{stform-bis})}. Clearly, the L\'evy-Kintchine formula
outlined in Sect.~{\ref{basic}} applies to translation-invariant
Markovian semigroup $\{\prop_t\}_{t\in\errep}$ (with $G=\erre^n$, of
course), and the hypothesis that, for all $\epsilon
>0 $, $ \prm_t (\{y\colon |y|\ge\epsilon\})= \ot$, for $t\downarrow
0$, implies that the L\'evy measure $\eta$ appearing
in~{(\ref{interm})} is identically zero (as a consequence of
relation~{(\ref{caralevi})}). Therefore,
formula~{(\ref{stform-bis})} is coherent with the L\'evy-Kintchine
formula~{(\ref{LKF})} (with $\rest\equiv 0$).

Finally, what we have recalled about the one-dimensional Brownian
motion is easily recovered as a particular case. Let
$\{\prop_t\}_{t\in\errep}$ be a translation-invariant Markovian
semigroup in $\cz(\erre)$ such that the associated convolution
semigroup of measures $\{\mut\}_{t\in\errep}$
satisfies~{(\ref{supponi})}. Then, its infinitesimal generator
$\infgen$ is uniquely determined by
\begin{equation}
\big(\infgen\tre f\big)(x)=  b\cinque \frac{\partial}{\partial
x}\tre f(x) +  a\cinque \frac{\partial^2}{\partial x^2}\tre f(x),\ \
\ \forall\quattro f\in \cdc(\erre),
\end{equation}
for some $a,b\in\erre$, $a\ge 0$. If $a>0$, for every $f\in
\con(\erred)$, we have that
\begin{equation}
\big(\prop_t\tre f\big)(x)=\int_{\erre} f(y)\sette\wp_t(y-x)\; \de
y,\ \ \ t>0,
\end{equation}
where $\wp_{(\cdot)}(\cdot)\colon\erreps\times\erre\rightarrow\erre$
is the well known fundamental solution of the drift-diffusion
equation:\footnote{Namely, $\wp_t(y)=\frac{1}{\sqrt{4\pi at}}\cinque
\exp(-(y-bt)^2/4at)$, for $t>0$. }
\begin{equation}
\frac{\partial}{\partial t}\sette \wp_t(y) = -b\cinque
\frac{\partial}{\partial y}\sette \wp_t(y)  +  a\cinque
\frac{\partial^2}{\partial y^2}\sette \wp_t(y),\ \ \ t>0,\ a>0,\
b\in\erre.
\end{equation}
On the other hand, for $a=0$ we have a `pure drift regime' and $\mut
=\delta_{bt}$ (i.e., $|b|$ is the modulus of the drift velocity).
Suppose, now, that the semigroup $\{\prop_t\}_{t\in\errep}$ commutes
with the reflection $x\mapsto -x$ as well. Then, it follows that
$b=0$. Moreover, if $a> 0$ (standard Brownian regime), the
probability measure $\mut$, for $t>0$, is absolutely continuous with
respect to the Lebesgue measure on $\erre$  and the Radon-Nikodym
derivative $\wp_t$ of $\mut$ with respect to this measure satisfies
the diffusion equation~{(\ref{diffusion})}, with $D=a$. Otherwise
($a=0$), $\sopa=0$ and $\mut =\delta$, for all $t\in\errep$.

%--------------------------------------------------------------------------------
\section{Twirling superoperators and twirling semigroups}
\label{twirling}
%--------------------------------------------------------------------------------

In Sects.~{\ref{basic}} and~{\ref{errenne}},
we have introduced the notion of left-invariant
Markovian semigroup of operators in the Banach space
$\cz(G)$, with $G$ denoting a l.c.s.c.\ group, and we
have illustrated this notion in the remarkable case where
$G=\erre^n$. In this section, we will consider a class of semigroups
of operators that is the central object of the paper. More precisely,
we will deal with semigroups of `superoperators' acting in Banach spaces
of operators. The most evident link between the two mentioned classes of
operator semigroups is given by the fact that both are defined by means of
convolution semigroups of probability measures on groups.

For the sake of clarity, we will establish the following notation.
Given a (separable complex) Hilbert space $\mathcal{H}$, we will
denote by $\opb$ a generic linear operator belonging to the Banach
space $\bH$ of bounded operators in $\hh$. The symbols $\opa$,
$\ops$ will denote generic operators in $\trc$ --- the Banach space
of trace class operators, endowed with the trace norm $\|\cdot\trn$
--- and in the Hilbert-Schmidt space $\hs$ (endowed with the norm
$\|\cdot\norhs$ induced by the Hilbert-Schmidt scalar product),
respectively. As is well known, $\trc$ and $\hs$ are two-sided
ideals in $\bH$, and $\trc\subset\hs$. We will denote by $\supops$,
$\dsupops$ the Banach spaces of bounded (super)operators in $\trc$
and $\bH$, respectively.

Let $G$ be a l.c.s.c.\ group, and let $U$ be a projective
representation of $G$ in $\mathcal{H}$. The following facts will be
very useful for our purposes. The map
\begin{equation}
\rep \colon G\rightarrow\mathcal{U}(\hs),
\end{equation}
defined by
\begin{equation} \label{definrep}
\rep(g)\hspace{0.3mm} \ops := U(g)\, \ops\, U(g)^\ast, \ \ \ \forall
\hspace{0.5mm}g\in G,\ \ \forall
\hspace{0.5mm}\ops\in\hs ,
\end{equation}
is a strongly continuous \emph{unitary} representation, even in the
case where the representation $U$ is genuinely \emph{projective};
see~{\cite{AnielloSP}}. Clearly, for every $g\in G$ the unitary
operator $\rep(g)$ in $\hs$ induces the Banach space isomorphism (a
surjective isometry) $\trc\ni\opa\mapsto\rep(g)\tre \opa\in\trc$.
Therefore, we can define the isometric representation
\begin{equation}
\urep\colon G \rightarrow\supops,\ \ \ \urep(g)\hspace{0.3mm} \opa
:= U(g)\, \opa\, U(g)^\ast, \ \ \ \forall \hspace{0.5mm}g\in G,\ \
\forall \hspace{0.5mm}\opa\in\trc,
\end{equation}
and it is obvious that $\urep(g)\hspace{0.3mm} \opa=\rep(g)\tre
\opa$, for all $\opa\in\trc$ and $g\in G$.
\begin{proposition} \label{conturep}
The isometric representation $\urep$ of the l.c.s.c.\ group $G$ in
the Banach space $\trc$ is strongly continuous.
\end{proposition}

\noindent {\bf Proof:} Since $G$ is a second countable (\emph{a
fortiori}, first countable) topological space, it is sufficient to
show that $\urep$ is sequentially continuous. Let
$\{g_n\}_{n\in\mathbb{N}}$ be a sequence in $G$ converging to $g$.
Then, for every $\opa\in\trc$, the sequences
\begin{equation}
\big\{\urep(g_n)\tre\opa=\rep(g_n)\tre\opa\big\}_{n\in\mathbb{N}},\
\ \
\big\{\big(\urep(g_n)\tre\opa\big)^\ast=\rep(g_n)\tre\opa^\ast\big\}_{n\in\mathbb{N}},
\end{equation}
converge to $\urep(g)\tre\opa$ and $\urep(g)\tre\opa^\ast$,
respectively, with respect to the Hilbert-Schmidt norm (the unitary
representation $\rep$ is strongly continuous), hence, with respect
to the strong operator topology in $\bH$. Applying `Gr\"umm's
convergence theorem' (see~{\cite{Simon-book}}, Chapter~2), by this
fact and by the fact that the representation $\urep$ is isometric,
we find out that the sequence
$\big\{\urep(g_n)\tre\opa\big\}_{n\in\mathbb{N}}$ converges to
$\urep(g)\tre\opa$ with respect to the trace norm, as
well.~{$\square$}

Next, observe that, for every $\opb\in\bH$, the map $G\ni g\mapsto
U(g)^\ast\tre \opb\sei U(g)\in\bH$ is weakly continuous
($\langle\phi, U(g)^\ast\tre \opb\sei
U(g)\sei\psi\rangle=\tr\big(\opb\big(\urep(g)\tre|\psi\rangle\langle\phi|\big)\big)$,
for all $\phi,\psi\in\hh$). Then, given a \emph{finite} Borel
measure $\mu$ on $G$, one can consider the bounded linear map
$\dwi\colon \bH\rightarrow\bH$ defined by
\begin{equation} \label{defidwi}
\dwi \opb := \int_G \de\mu(g)\ U(g)^\ast\tre \opb\sei U(g) ,\ \ \
\opb\in\bH,
\end{equation}
where on the r.h.s.\ of~{(\ref{defidwi})} a weak integral (i.e., an
integral converging with respect to the weak operator topology in
$\bH$) is understood. In the case where $\mu$ is normalized
($\mu(G)=1$; i.e., $\mu$ is a probability measure), it is obvious
that $\dwi I= I$ and it is easy to check that the linear map $\dwi$
is a contraction (i.e.\ its norm is not larger than one). From this
point onwards, we will assume that $\mu$ belongs to $\smis$.

It is clear that the map $\dwi$ is positive. One can prove,
moreover, that it is \emph{completely} positive. In fact, recalling
the necessary and sufficient condition~{(\ref{iff})}, for every
$m\in\mathbb{N}$ the positivity of the function $\mat\colon
G\rightarrow\erre$,
\begin{equation}
\mat (g) := \sum_{j,k=1}^{m}\big\langle\psi_j,
U(g)^\ast\hspace{0.2mm} \opb_j^\ast\tre\opb_k^{\phantom{\ast}}\sette
U(g)\cinque \psi_k\big\rangle,\ \ \ g\in G,
\end{equation}
for any $m$-tuple $\{\psi_1,\ldots ,\psi_m\}$ in $\hh$
and any $m$-tuple $\big\{\opb_1^{\phantom{\ast}},\ldots,\opb_m^{\phantom{\ast}}\big\}$
in $\bH$, implies that
\begin{equation}
\sum_{j,k=1}^{m}\big\langle\psi_j,\dwi \big(\opb_j^\ast\tre\opb_k^{\phantom{\ast}}
\big)\tre\psi_k\big\rangle=
\int_G \de\mu(g)\ \mat(g)\ge 0.
\end{equation}

As is well known, the dual space of $\trc$ can be identified with $\bH$
via the pairing
\begin{equation}
\bH\times\trc\ni (\opb,\opa) \mapsto \tr\big(\opb\opa\big)\ni\ccc.
\end{equation}
One can show that the map $\dwi$ is the adjoint of the linear
map $\twi\colon\trc\rightarrow\trc$ defined by
\begin{equation} \label{defitwi}
\twi \opa := \int_G \de\mu(g)\; \big(\urep(g)\tre \opa\big) ,\ \ \
\opa\in\trc,
\end{equation}
where, again, a weak integral (weak operator topology in $\bH$) is
understood. Observe, in fact, that $\twi\opa$ is a bounded operator
(and $\opa\ge 0 \Rightarrow \twi \opa\ge 0$); moreover, it is in the
trace class and
\begin{equation} \label{trpr}
\tr\big(\twi \opa\big) =\tr\big(\opa\big),\ \ \ \forall
\quattro\opa\in\trc.
\end{equation}
This last assertion is verified assuming --- without loss of
generality, since $\opa\in\trc$ can be expressed as a linear
combination of four \emph{positive} trace class
operators,\footnote{The positive operators $\opa_1,\ldots, \opa_4$ are
uniquely determined by the additional condition that $\opa_1\tre\opa_2=0=
\opa_3\tre\opa_4$. If this condition holds, then $\big\|\opa_1-\opa_2\big\trn=\tr\big(\opa_1\big) +
\tr\big(\opa_2\big)$.}
namely, $\opa = \opa_1-\opa_2 +\ima\big(\opa_3-\opa_4\big)$
--- that $\opa$ is positive, and using the definition of the trace and
the `monotone convergence theorem' for permuting the possibly
infinite sum (associated with the trace) with the integral on $G$.
Next, one can verify that
\begin{equation} \label{verrel}
\tr\big(\opb \big(\twi \opa\big)\big) =\tr\big(\big(\dwi
\opb\big)\opa\big),\ \ \ \forall \quattro\opa\in\trc,\ \forall
\quattro\opb\in\bH .
\end{equation}
To this aim, assume --- again, without loss of generality --- that
$\opa\in\trc$ and $\opb\in\bH$ are both positive. Then, given an
orthonormal basis $\{\psi_l\}_{l\in\den}$ in $\hh$
($\den\subset\mathbb{N}$), we have:
\begin{eqnarray}
\tr\big(\opb \big(\twi \opa\big)\big) \spa & = & \spa
\tr\big(\opb^{1/2} \big(\twi \opa\big)\opb^{1/2}\big)
\nonumber\\ \label{posint}
& = & \spa \sum_{l\in\den}\int_G \de\mu(g)\
\big\langle\psi_l,\opb^{1/2}\cinque U(g)\tre \opa\sei
U(g)^\ast\tre\opb^{1/2}\tre\psi_l\big\rangle .
\end{eqnarray}
At this point, since the integrand function on the r.h.s.\ of~{(\ref{posint})}
is positive, we can apply the `monotone convergence theorem'
and permute the
(possibly infinite) sum with the integral, thus getting
\begin{eqnarray}
\tr\big(\opb \big(\twi \opa\big)\big) \spa & = & \spa \int_G
\de\mu(g)\ \tr\big(\opb^{1/2}\cinque U(g)\tre \opa\sei
U(g)^\ast\tre\opb^{1/2}\big)
\nonumber\\
& = & \spa \int_G \de\mu(g)\ \tr\big(\opa^{1/2}\cinque U(g)^\ast\tre
\opb\sei U(g)\tre\opa^{1/2}\big)
\nonumber\\ \label{linea}
& = & \spa \int_G
\de\mu(g)\sum_{l\in\den}\big\langle\psi_l,\opa^{1/2}\cinque
U(g)^\ast\tre \opb\sei U(g)\tre\opa^{1/2}\tre\psi_l\big\rangle .
\end{eqnarray}
Eventually, we can again permute the sum with the integral and
obtain relation~{(\ref{verrel})}. Note that the first line of
~{(\ref{linea})} implies that $\twi$  coincides with the weak
integral --- i.e., the integral with respect to the weak topology of
bounded operators in $\trc$ --- $\int_G \de\mu(g)\; \urep(g)$. Also
note that, since $\dwi$ is a contraction in $\bH$, $\twi$ is a
contraction in $\trc$; indeed:
\begin{eqnarray}
\big\|\twi\opa\big\trn  \spa & = & \spa \sup\big\{
\big|\tr\big(\opb\big(\twi\opa\big)\big)\big|\colon\cinque
\opb\in\bH,\ \|\opb\|=1\big\}
\nonumber\\
& = & \spa \sup\big\{ \big|\tr\big(\big(\dwi
\opb\big)\opa\big)\big|\colon\cinque \opb\in\bH,\ \|\opb\|=1\big\}
\nonumber\\
& \le & \spa \|\opa\trn\cinque \sup\big\{ \big\|\dwi
\opb\big\|\colon\cinque \opb\in\bH,\ \|\opb\|=1\big\} \le
\|\opa\trn,
\end{eqnarray}
for all $\opa\in\trc$.

We can summarize our previous discussion by stating the following result.
\begin{proposition}
For every projective representation $U$ of a l.c.s.c.\ group $G$ in
$\hh$ and for every probability measure $\mu$ on $G$, the bounded
linear map $\twi\colon\trc\rightarrow\trc$ defined
by~{(\ref{defitwi})} is a contraction, and it is positive and
trace-preserving. Moreover, we have the formula
\begin{equation} \label{concisa}
\twi = \int_G \de\mu(g)\ \urep(g),
\end{equation}
where the integral holds in the weak sense. The bounded linear map
$\dwi\colon \bH\rightarrow\bH$ defined by~{(\ref{defidwi})} is the
adjoint of $\twi$. It is a completely positive map.
\end{proposition}

\begin{remark} \label{bistochastic}
{\rm Suppose that the Hilbert space of the representation $U$ is
finite-dimensional. Then, for every probability measure $\mu$ on
$G$, $\twi$ is a completely positive, trace-preserving linear map
which is also \emph{unital}, i.e., such that $\twi I=I$. Therefore,
it is a \emph{bistochastic} (or `doubly stochastic') linear
map~{\cite{Bengtsson}}. Clearly, the bistochastic linear maps in
$\supops$ form a convex set. The determination of the extreme points
of this convex set is an interesting problem~{\cite{Streater}}. From
the physicist's point of view, these maps are characterized by the
property of leaving the maximally mixed state
invariant.~{$\blacksquare$} }
\end{remark}

In the case where $G$ is a unitary group ($\mathrm{U}(n)$ or
$\mathrm{SU}(n)$), $\mu$ is the Haar measure on $G$ (normalized in
such a way that $\mu(G)$=1) and $U$ is the defining representation
of $G$, we have that $\twi$ is the `standard' \emph{twirling
superoperator} (in $\mathcal{B}(\ccc^n)$). Therefore, in the general
case, it is quite natural to extend this terminology and call $\twi$
the \emph{$(U,\mu)$-twirling superoperator}; the map $\dwi$ will be
called, accordingly, the \emph{dual $(U,\mu)$-twirling
superoperator}. Since any convex combination of two probability
measures on $G$ is again a probability measure, the following result
holds.

\begin{proposition}
For every projective representation $U$ of $G$ in $\hh$, the subsets
$\big\{\twi\colon\; \mu\in\smis\big\}$, $\big\{\dwi\colon\;
\mu\in\smis\big\}$ of the Banach spaces $\supops$ and $\dsupops$,
respectively, are convex.
\end{proposition}

\begin{remark} \label{Bochner}
{\rm It is worth observing that in definition~{(\ref{defitwi})} of
the twirling superoperator one may replace the weak integral with a
Bochner integral (relative to the Banach space $\trc$).

It is also an interesting fact that a probability measure $\mu$ on
$G$ allows us to define a bounded linear map
$\twihs\colon\hs\rightarrow\hs$ along the scheme already outlined
for the maps $\dwi$ and $\twi$, i.e.,
\begin{equation} \label{defitwihs}
\twihs \ops := \int_G \de\mu(g)\; \big(\rep(g)\tre \ops\big) ,\ \ \
\ops\in\hs,
\end{equation}
where, once again, one can show that the map $\twihs$ is well
defined (with the integral on the r.h.s.\ of~{(\ref{defitwihs})}
regarded, equivalently, as a weak or as a Bochner integral). Indeed,
observe that, for every $\ops\in\hs$, we have:
\begin{eqnarray}
0 \spa & < & \spa \sum_{l\in\den}\sei \int_G \de\mu(g) \int_G
\de\mu(h)\ \langle \psi_l, U(g)\tre \ops^\ast\tre U(g)^\ast\cinque
U(h)\tre \ops\sei U(h)^\ast\psi_l\rangle \nonumber\\
& \le & \spa \sum_{l\in\den}\sei \int_G \de\mu(g) \int_G \de\mu(h)\
|\langle \psi_l, U(g)\tre \ops^\ast\tre U(g)^\ast\cinque U(h)\tre
\ops\sei U(h)^\ast\psi_l\rangle|
\nonumber\\
& \le & \spa  \int_G \de\mu(g) \int_G \de\mu(h)\quattro
\sum_{l\in\den}\sei |\langle \psi_l, U(g)\tre \ops^\ast\tre
U(g)^\ast\cinque U(h)\tre \ops\sei U(h)^\ast\psi_l\rangle|
\nonumber\\
& \le & \spa  \int_G \de\mu(g) \int_G \de\mu(h)\hspace{1mm}
\|U(g)\tre \ops^\ast\tre U(g)^\ast\cinque U(h)\tre \ops\sei
U(h)^\ast\trn \le \|\ops\norhssq^2.
\end{eqnarray}
The previous argument also shows that $\twihs$ is a contraction. It
is clear, moreover, that the map $\twi$ can be regarded as the
restriction to the trace class operators of the map
$\twihs$.~{$\blacksquare$} }
\end{remark}

From definition~{(\ref{defitwi})} it is clear that the map
\begin{equation} \label{homo}
\smis\ni\mu\mapsto\twi\in\cpt
\end{equation}
is a homomorphism of the semigroup $\smis$ --- with respect to
convolution --- into the semigroup $\cpt$ --- with respect to
composition --- of (quantum) \emph{dynamical maps} in $\trc$,
namely, of the semigroup consisting of all positive,
trace-preserving, bounded linear maps in $\trc$, whose adjoints
(acting in the Banach space $\bH$) are completely
positive~{\cite{Holevo}}. This observation leads us to consider an
interesting class of continuous one-parameter semigroups of
superoperators.

Indeed --- given  a continuous one-parameter convolution semigroup
$\{\mut\}_{t\in\errep}\subset\smis$ of (probability) measures on $G$
and a projective representation $U$ of $G$ in $\hh$
--- for every $t\ge 0$, we can as above define the $(U,\mut)$-twirling superoperator
\begin{equation}
\twis\equiv \twit\colon \trc\rightarrow \trc,\ t\ge 0,\ \ \
(\mathfrak{S}_0 =I).
\end{equation}
The fact that $\{\twis\}_{t\in\errep}$ enjoys the one-parameter
semigroup property is a consequence of the fact that
$\{\mut\}_{t\in\errep}$ is a convolution semigroup and the
map~{(\ref{homo})} is a homomorphism. Moreover, the semigroup
$\{\twis\}_{t\in\errep}$ is strongly right continuous at $t=0$. This
is a consequence of the continuity of $\{\mut\}_{t\in\errep}$ and of
Proposition~{\ref{conturep}}. Actually, as recalled in
Sect.~{\ref{basic}}, it suffices to prove the weak right continuity
at $t=0$ of the semigroup $\{\twis\}_{t\in\errep}$. To this aim,
observe that, for every $\opa\in\trc$ and $\opb\in\bH$, the function
\begin{equation} \label{confu}
G\ni g \mapsto\tr\big(\opb\tre
\big(\urep(g)\tre\opa\big)\big)\in\ccc
\end{equation}
is continuous (equivalently, the representation $\urep$ is weakly
continuous). Also note that
\begin{equation}
\big |\tr\big(\opb \tre \big(\urep(g)\tre\opa\big)\big)
-\tr\big(\opb \tre \opa\big) \big |\le \big\|\opb \tre
\big(\urep(g)\tre\opa\big)\trn + \big\|\opb \tre \opa \big\trn \le
2\quattro \big\|\opb\big\| \quattro\big\| \tre \opa \big\trn,
\end{equation}
for all $g\in G$. Therefore, the function
\begin{equation}
G\ni g\mapsto\big |\tr\big(\opb \tre \big(\urep(g)\tre\opa\big)\big)
-\tr\big(\opb \tre \opa\big)\big |\in\erre
\end{equation}
is bounded and continuous. At this point, we can exploit the fact
that $\lim_{t\downarrow 0} \mut = \delta$ (weakly). By this
relation, since
\begin{eqnarray}
\big |\tr\big(\opb \big(\twis \opa\big)\big)-\tr\big(\opb \tre
\opa\big)\big | \spa & = & \spa \Big | \int_G \de\mut(g)\;\Big(
\tr\big(\opb \tre \big(\urep(g)\tre\opa\big)\big)
-\tr\big(\opb \tre \opa\big) \Big)\Big | \nonumber\\
& \le & \spa \int_G \de\mut(g)\ \big |\tr\big(\opb \tre
\big(\urep(g)\tre\opa\big)\big) -\tr\big(\opb \tre \opa\big)\big |,
\end{eqnarray}
we conclude that
\begin{equation}
\lim_{t\downarrow 0} \big |\tr\big(\opb \big(\twis
\opa\big)\big)-\tr\big(\opb \tre \opa\big)\big | = 0,\ \ \
\forall\quattro\opa\in\trc,\ \forall\quattro\opb\in\bH.
\end{equation}
This completes the proof of the continuity of the one-parameter
semigroup $\{\twis\}_{t\in\errep}$.

At this point, recalling that a \emph{quantum dynamical
semigroup}~{\cite{Holevo}} in $\trc$ is a (strongly) continuous
one-parameter semigroup of quantum dynamical maps in $\trc$, we can
resume our preceding discussion stating the following result.
\begin{proposition}
The contraction semigroup
$\{\twis\colon\trc\rightarrow\trc\}_{t\in\errep}$ is a quantum
dynamical semigroup.
\end{proposition}

\begin{remark}
{\rm Recalling Remark~{\ref{bistochastic}}, we have that --- in the
case where the Hilbert space of the representation $U$ is
finite-dimensional --- the dynamical semigroup
$\{\twis\}_{t\in\errep}$ is a \emph{bistochastic dynamical
semigroup}. A complete characterization of the twirling semigroups
associated with finite-dimensional representations of Lie groups
will be provided in Sect.~{\ref{groups}}.~{$\blacksquare$}}
\end{remark}

\begin{remark}
{\rm The contraction $\twihs$ defined by~{(\ref{defitwihs})} allows
us to define, for every continuous convolution semigroup
$\{\mut\}_{t\in\errep}$ of probability measures on $G$, a
contraction semigroup $\{\twishs\}_{t\in\errep}$ in the Hilbert
space $\hs$, i.e.
\begin{equation}
\twishs\equiv \twiths\colon \hs\rightarrow \hs.
\end{equation}
The fact that $\wlim_{t\downarrow 0}\twishs=I$ can be proved by
means of a procedure analogous to that adopted for the semigroup
$\{\twis\}_{t\in\errep}$.~{$\blacksquare$}}
\end{remark}

In the following, we will call $\{\twis\}_{t\in\errep}$ the
\emph{twirling semigroup} associated with (or induced by) the pair
$(U,\{\mut\}_{t\in\errep})$. We stress that, in general, a twirling
semigroup will be induced by different pairs of the type (projective
representation, convolution semigroup of measures).

%------------------------------------------------------------------------------
\section{Brownian motion on Lie groups and open quantum systems}
\label{groups}
%------------------------------------------------------------------------------

In this section, we will study the twirling semigroups of operators
induced by representations of \emph{Lie groups}. This is a
particularly interesting case because the differential structure of
a Lie group allows us to obtain a characterization of the
infinitesimal generators of the associated twirling semigroups. The
main technical tool will be the L\'evy-Kintchine
formula~{(\ref{LKF})}. In order to avoid all mathematical
intricacies related to infinite-dimensional Hilbert spaces, we will
consider the case where the group representations involved are
finite-dimensional, case which is relevant, for instance, in
applications to quantum computation~{\cite{Nielsen}}. The general
case will be considered elsewhere.

Thus, in the following we will deal with a smooth, finite-dimensional
unitary representation $U$ of a Lie group $G$ (of dimension $n$) in
a $\dime$-dimensional (complex) Hilbert space $\hh$. It is clear
that, in this case, $\bH=\trc=\hs$ and $\supops=\dsupops$. Since all
norms in $\bH$ (or $\supops$) induce the same topology (as $\hh$ is
finite-dimensional), all our statements involving topological
properties of $\bH$ (or $\supops$) --- convergence, continuity,
compactness \emph{et cetera} --- are to be understood as referred to
this topology. We will denote, as usual, by
$\mathcal{U}(\mathcal{H})$ the unitary group of $\hh$, endowed with
the topology inherited from $\bH$; it is well known that
$\mathcal{U}(\mathcal{H})$ is compact with respect to this topology.
Let us fix once and for all a basis $\{\xi_1,\ldots,\xi_n\}$ in the
Lie algebra $\lie$ and a system of adapted coordinates $\{g\mapsto
\xe^1(g), \ldots, g\mapsto \xe^n(g)\}$ based at the identity. We
will use the notations adopted in Sect.~{\ref{basic}}, usually with
no further explanation.

\begin{remark} \label{inteco}
{\rm We will repeatedly use the following fact. Let $f\colon
G\rightarrow\supops$ a bounded continuous function. Then, for every
probability measure $\mu$ on $G$, $\mu(f):=\int_G f(g)\, \de\mu(g)$
belongs to the closure of the convex hull $\co(f(G))\subset\supops$.
Indeed, observe that $G$ is (homeomorphic to) a separable metric
space. Then, there exists a sequence $\{\mu_m\}_{m\in\mathbb{N}}$ of
\emph{finitely supported} probability measures on $G$ weakly
converging to $\mu$ (see~{\cite{Parthasarathy}}, chapter~2,
Theorem~6.3). Hence, $\mu_m(f)\in\co(f(G))$ and
$\mu(f)=\lim_{m\rightarrow\infty}
\mu_m(f)\in\clco(f(G))$.~{$\blacksquare$}}
\end{remark}

We have observed in Sect.~{\ref{twirling}} that a twirling superoperator is
a bistochastic linear map, see Remark~{\ref{bistochastic}}.
We will now show that, actually, it belongs to a special class of bistochastic maps,
namely, the class of `random unitary maps'.

\begin{definition}
A quantum dynamical map $\randu\colon\bH\rightarrow\bH$ is said to
be a \emph{random unitary map} if it admits a decomposition of the
form
\begin{equation} \label{defirandu}
\randu\sette \opa= \sum_{k=1}^{\den} p_k\tre \vk\opa\otto\vkast,\ \ \ \den\in\mathbb{N},
\end{equation}
where  $\{\vk\}_{k=1}^{\den}$ is a set of unitary operators in $\hh$
and $\{p_k\}_{k=1}^{\den}\subset\erreps$ is a probability
distribution; i.e., if it is a convex combination of unitary
transformations. The \emph{cardinality} $\cardi(\randu)$ of a random
unitary map $\randu$ is the minimum number of terms required in a
decomposition of $\randu$ of the type~{(\ref{defirandu})}.
\end{definition}
Observe that the random unitary maps acting in $\bH$ form a
semigroup $\cptru$ contained in the semigroup of quantum dynamical
maps $\cpt$. It is natural to consider the nonzero positive integer
$\cardin$ defined as follows:
\begin{equation}
\cardin:=\sup \big\{\cardi(\randu)\in\mathbb{N}\colon
\randu\in\cptru \big\},\ \ \ \dime=\dim(\hh).
\end{equation}
Since a random unitary map sends the subspace, formed by the traceless operators,
of the real vector space $\brH$ (of
selfadjoint operators in $\hh$)
into itself, applying Carath\'eodory
theorem one finds the estimate $\cardin\le \left(\dime^2-1\right)^2 +1= \dime^4-2\tre\dime^2+2$.
This estimate is not tight. For instance, in the case where $\dime=2$,
it is known that all bistochastic maps (hence, all random unitary maps)
are `Pauli channels'~{\cite{Bengtsson}}; thus, $\mathsf{c}(2)=4$. To the best of
our knowledge, the generic integer $\cardin$ is unknown, but stricter bounds for the cardinality
of a random unitary map can be provided and it turns out that $\cardin\le \dime^2$~{\cite{Buscemi}}.

Consider, now, a subgroup $\suni$ of the group $\uni(\hh)$. The
closure $\csuni$ of $\suni$ is a subgroup of $\uni(\hh)$, as well.
Denote by $\dmru(\suni)$ the subset of $\cptru$ formed by those
superoperators of the form~{(\ref{defirandu})} with the set of
unitary operators $\{\vk\}_{k=1}^{\den}$ contained in $\suni$.
Clearly, $\cptru =\dmru(\uni(\hh))$, and $\dmru(\suni)$ is a
subsemigroup of $\cptru$. It is clear that, defining
\begin{equation} \label{defvvvv}
\vvv := \big\{V\tre(\cdot)\tre V^\ast\in\supops\colon\;
V\in\suni\big\},
\end{equation}
the semigroup $\dmru(\suni)$ is nothing but the convex hull of
the set $\vvv$:
\begin{equation}
\dmru(\suni)=\co(\vvv).
\end{equation}

\begin{lemma} \label{lemmaran}
For every subgroup $\suni$ of $\uni(\hh)$, the semigroup
$\dmru(\csuni)$ is a compact convex subset of $\supops$ that
coincides with the set $\overline{\dmru(\suni)}$. Thus, in
particular, the semigroup $\cptru$ is a compact convex subset of
$\supops$.
\end{lemma}

\noindent {\bf Proof:} Note that the map
\begin{equation} \label{vvmap}
\mathcal{U}(\mathcal{H})\ni V\mapsto V\tre(\cdot)\tre V^\ast\in\supops
\end{equation}
is continuous. Hence, the image, through this map, of the closed
subgroup $\csuni$ of $\uni(\hh)$ --- i.e.\ $\cvvv$ --- is a compact subset $\mathcal{K}$
of $\supops$. Recall that, in a finite-dimensional (real or complex)
vector space, the convex hull of a compact set of is compact, and
the closure of the convex hull of a bounded set coincides with the
convex hull of the closure of this set. Then,
$\dmru(\csuni)=\co(\cvvv)$ is a compact subset of $\supops$.
Moreover, $\co(\cvvv)$ coincides with the closure $\clco(\vvv)=\overline{\dmru(\suni)}$ of
$\co(\vvv)$. Indeed, $\cvvv=\overline{\vvv}$
(as the map~{(\ref{vvmap})} is
continuous, $\cvvv\subset\overline{\vvv}$, and $\cvvv=\overline{\cvvv}\supset\overline{\vvv}$);
hence: $\co(\cvvv)=\co(\overline{\vvv})=\clco(\vvv)$.~{$\square$}

\begin{definition}
A \emph{random unitary semigroup} acting in $\bH$ is a quantum dynamical semigroup taking values
in the semigroup $\cptru$.
\end{definition}

\begin{proposition} \label{twi-ran}
Every twirling superoperator in $\bH$ is a random unitary map. Therefore,
every twirling semigroup acting in $\bH$ is a random unitary semigroup.
\end{proposition}

\noindent {\bf Proof:} The expression~{(\ref{concisa})} of a twirling superoperator
involves an integral that, in the case where $\hh$ is finite-dimensional, can be considered
to be defined with respect to the topology of $\supops$.
Thus, taking into account Remark~{\ref{inteco}},
from Lemma~{\ref{lemmaran}} the statement follows.~{$\square$}

A quantum dynamical semigroup $\{\qds\colon
\bH\rightarrow\bH\}_{t\in\errep}$ is completely characterized by its
(in this case, of course, bounded) infinitesimal generator $\gene$:
\begin{equation} \label{QDSG}
\gene = \lim_{t\downarrow 0}
t^{-1}\big(\qds -I).
\end{equation}
According to the Gorini-Kossakowski-Lindblad-Sudarshan classification
theorem~{\cite{Gorini,Lindblad}},
$\gene$ has the general form
\begin{equation} \label{forgene}
\gene\sei\opa = - \ima \big[\oph, \opa\big] +
\cpm\sei\opa - \frac{1}{2}
\left((\cpm^\ast I)\tre\opa+\opa\sette(\cpm^\ast I)\right),
\end{equation}
where $\oph$ is a trace-less selfadjoint operator in $\hh$,
$\cpm \colon \bH\rightarrow\bH$ a completely positive map and
$\cpm^\ast$ its adjoint with respect to the Hilbert-Schmidt scalar product in $\bH$.
\begin{remark}
{\rm As is well known~{\cite{Bengtsson}}, a completely positive map
$\mathfrak{K}\colon \bH\rightarrow\bH$ can be expressed in the
\emph{Kraus-Stinespring-Sudarshan canonical form}:
\begin{equation}
\mathfrak{K} \big(\opa\big) = \sum_{k=1}^{\dime^2}
\gamma_k\sei\opk_k^{\phantom{\ast}}\opa\sei\opk_{k}^\ast,\ \ \
\gamma_k\ge 0,\ \ \ \opa\in\bH,
\end{equation}
where $\opk_1^{\phantom{\ast}},\ldots,
\opk_{\dime^2}^{\phantom{\ast}}$ are linear operators in $\hh$ such
that
\begin{equation}
\big\langle
\opk_{j}^{\phantom{\ast}},\opk_k^{\phantom{\ast}}\big\ranglehs :=
\tr\big(\opk_{j}^\ast\cinque\opk_k^{\phantom{\ast}}\big)=\delta_{jk},\
\ \ j,k =1,\dots,\dime^2.
\end{equation}
However, it can be easily shown that the completely positive map
$\cpm$ in formula~{(\ref{forgene})} can be assumed, without loss of
generality, to be of the form
\begin{equation} \label{noloss}
\cpm\sei \opa = \sum_{k=1}^{\dime^2-1}
\gamma_k\sei\opf_k^{\phantom{\ast}}\opa\sei\opf_{k}^\ast,\ \ \
\gamma_k\ge 0,\ \ \ \ \ \left(\cpm^\ast \opa =
\sum_{k=1}^{\dime^2-1}
\gamma_k\sei\opf_{k}^\ast\opa\sei\opf_k^{\phantom{\ast}}\right),
\end{equation}
where the $\dime^2-1$ linear operators
$\opf_1^{\phantom{\ast}},\ldots, \opf_{\dime^2-1}^{\phantom{\ast}}$
form an orthonormal basis --- with respect to the Hilbert-Schmidt
scalar product $\langle\cdot,\cdot\ranglehs$ --- in the orthogonal
complement of the one-dimensional subspace of $\bH$ generated by the
identity operator (thus, they are trace-less). In this way, formula~{(\ref{forgene})}
gives the so-called `diagonal form'~{\cite{Breuer}} of the infinitesimal
generator $\gene$.~{$\square$} }
\end{remark}

Later on, we will prove a generalization of a classical result  of
K\"ummerer and Maassen~{\cite{Kummerer}}; see
Theorem~{\ref{main-th}} below. As a first step, from
ref.~{\cite{Kummerer}}  we can extract some useful information on
random unitary semigroups. Given a subgroup $\suni$ of the group
$\uni(\hh)$, we will denote by $\coco(\suni)$ the closure of the
convex cone in $\supops$ generated by the set $\vvv -I$; namely,
\begin{equation} \label{vvset}
\coco(\suni):=\clcocone\big(\big\{(V\tre(\cdot)\tre
V^\ast-I)\in\supops\colon\; V\in\suni\big\}\big).
\end{equation}
In particular, we will adopt the shorthand notation $\coconeh\equiv\coco(\uni(\hh))$.
\begin{proposition} \label{propequi}
The following facts are
equivalent.
\begin{description}
\item[(a)]\hspace{0mm} The quantum dynamical semigroup $\{\qds\colon
\bH\rightarrow\bH\}_{t\in\errep}$ is a random unitary semigroup.
\item[(b)]\hspace{0mm} The infinitesimal generator of the quantum dynamical semigroup $\{\qds\colon
\bH\rightarrow\bH\}_{t\in\errep}$ belongs to the closed convex cone $\coconeh$.
\item[(c)]\hspace{0mm} The infinitesimal generator $\gene$ of the quantum dynamical semigroup $\{\qds\colon
\bH\rightarrow\bH\}_{t\in\errep}$ is of the form~{(\ref{forgene})},
with the completely positive map $\cpm \colon \bH\rightarrow\bH$ of
the form
\begin{equation}
\cpm\sei \opa =  \sum_{k=1}^{\mathtt{K}}
\ope_k^{\phantom{\ast}}\opa\sei\ope_{k}+\gamma_0\cinque\randu\sette
\opa,\ \ \ \ope_k^{\phantom{\ast}}\in\brH,\ \ \ \gamma_0\ge 0,\ \ \
\randu\in\cptru,
\end{equation}
for all $\opa\in\bH$.
\item[(d)]\hspace{0mm} The infinitesimal generator $\gene$ of the quantum dynamical semigroup $\{\qds\colon
\bH\rightarrow\bH\}_{t\in\errep}$ is of the form
\begin{equation} %\label{forgeneg}
\gene\sei\opa = - \ima \big[\oph, \opa\big] + \sum_{k=1}^{\dime^2-1}
\gamma_k\Big(\opl_k^{\phantom{\ast}}\opa\sei\opl_{k}^{\phantom{\ast}}-\frac{1}{2}
\big(\opl_k^2\tre\opa+ \opa\sei\opl_k^2\big)\Big) \msei +
\gamma_0\big(\randu -I\big)\tre \opa,\ \ \ \opa\in\bH,
\end{equation}
where $\oph$ is a trace-less selfadjoint operator,
$\opl_1^{\phantom{\ast}},\ldots, \opl_{\dime^2-1}^{\phantom{\ast}}$
are trace-less selfadjoint operators such that
\begin{equation} \label{orthog}
\big\langle
\opl_{j}^{\phantom{\ast}},\opl_k^{\phantom{\ast}}\big\ranglehs
=\delta_{jk},\ \ \ j,k =1,\dots,\dime^2-1,
\end{equation}
$\randu$ is a random unitary map acting in $\bH$ and
$\gamma_0,\ldots,\gamma_{\dime^2-1}$ are non-negative numbers.
\end{description}
\end{proposition}

\noindent {\bf Proof:} The equivalence of (a), (b) and~{(c)} is
proved in~{\cite{Kummerer}} (see Theorem~{1.1.1.}; here we have only
adapted terminology and results to our context). The equivalence of
(c) and~{(d)} is straightforward. Hint: in order to get~{(d)}
from~{(c)}, expand the selfadjoint operators
$\big\{\ope_k^{\phantom{\ast}}\big\}_{k=1}^{\mathtt{K}}$
--- $\ope_k^{\phantom{\ast}}=\sum_{l=0}^{\dime^2-1} c_{kl}\tre
\opf_l^{\phantom{\ast}}$ --- with respect to an orthonormal basis
$\big\{\opf_l^{\phantom{\ast}}\mcinque\big\}_{l=0}^{\dime^2-1}$ in
$\brH$
($\big\langle\opf_{j}^{\phantom{\ast}},\opf_l^{\phantom{\ast}}\mcinque\big\ranglehs
=\delta_{jl}$), with $\opf_{0}^{\phantom{\ast}} = I$; then,
diagonalize the positive real matrix
$[\matri_{lm}]_{l,m=1}^{\dime^2-1}$, where $\matri_{lm}=
\sum_{k=1}^{\mathtt{K}} c_{kl}\cinque c_{km}$, by means of an
orthogonal transformation, and next use the orthogonal matrix
involved in this transformation for defining a new orthonormal basis
in the subspace of $\brH$ formed by the traceless
operators.~{$\square$}

For reasons that will be clear later on, it is convenient to
single out a special class of random unitary semigroups, namely, the
\emph{Gaussian dynamical semigroups}.
\begin{definition}
We will say that a quantum dynamical semigroup
$\{\qds\}_{t\in\errep}$ acting in $\bH$ is a \emph{Gaussian
dynamical semigroup} if its infinitesimal generator $\geneg$ can be
expressed in the form
\begin{equation} \label{forgeneg}
\geneg\sei\opa = - \ima \big[\oph, \opa\big] +
\sum_{k=1}^{\dime^2-1}
\gamma_k\Big(\opf_k^{\phantom{\ast}}\opa\sei\opf_{k}^{\phantom{\ast}}-\frac{1}{2}
\big(\opf_k^2\tre\opa+ \opa\sei\opf_k^2\big)\Big),
\end{equation}
where $\oph$ is a trace-less selfadjoint operator,
$\opf_1^{\phantom{\ast}},\ldots, \opf_{\dime^2-1}^{\phantom{\ast}}$
are trace-less selfadjoint operators satisfying~{(\ref{orthog})}
and
\begin{equation}
\gamma_1\ge0,\ldots, \gamma_{\dime^2-1}\ge 0,\ \ \ \gamma_1\tre\gamma_2\cdots\gamma_{\dime^2-1}\neq 0.
\end{equation}
\end{definition}
Otherwise stated, the infinitesimal generator $\gene$ of
formula~{(\ref{forgene})} gives rise to a Gaussian dynamical semigroup
if the completely positive map $\cpm$ admits a decomposition of the
form~{(\ref{noloss})} where the linear operators
$\opf_1^{\phantom{\ast}},\ldots, \opf_{\dime^2-1}^{\phantom{\ast}}$
are --- in addition to the previously mentioned assumptions ---
selfadjoint, and there is at least a nonzero number in the set
$\{\gamma_1,\ldots, \gamma_{\dime^2-1}\}$. Note that, according to
Proposition~{\ref{propequi}}, every Gaussian dynamical semigroup is
a random unitary semigroup. We will show, moreover, that every
Gaussian dynamical semigroup arises in a natural way as a twirling
semigroup associated with a convolution semigroup of measures of a certain type,
namely, with a `Gaussian semigroup of measures'.

In order to define such a class of convolution semigroups of measures,
let us consider the following set of probability measures on the Lie
group $G$:
\begin{equation}
\misd := \{\delta_g\colon g\in G\}\subset\smis;
\end{equation}
i.e., $\misd$ is the set of all Dirac measures on $G$.
\begin{definition}
A continuous convolution semigroup of measures $\{\mut\}_{t\in
\errep}$ --- such that, for $t>0$, $\mut\in\smis\smallsetminus\misd$
--- is called a \emph{Gaussian (convolution) semigroup of measures}
if
\begin{equation} \label{hypo}
\lim_{t\downarrow 0} t^{-1} \mut\big(\complement\tre\mathcal{E}_e\big)=0,
\end{equation}
for every Borel neighborhood of the identity $\mathcal{E}_e$ in $G$.
\end{definition}
The previous definition is originally due to
Courr\`ege~{\cite{Courrege}} and Siebert~{\cite{Siebert}}. Gaussian
semigroups of measures on $G$ describe the statistical properties of
Brownian motion on $G$~{\cite{Heyer}}. We have already encountered
condition~{(\ref{hypo})} --- see~{(\ref{supponi})} --- in the case
where $G=\erre^n$. Thus, the reader should be familiar with its
consequences. In general, it is a well known fact
--- see~{\cite{Heyer}}
--- that, given a Gaussian semigroup of measures $\{\mut\}_{t\in
\errep}$ on $G$, for every $t\in\errep$ the measure $\mut$ has
support contained in the connected component with the identity of
$G$: $\mathrm{supp}\tre(\mut)\subset G_e$. Therefore, in the
following we can assume without loss of generality that --- as far
as a Gaussian semigroup of measures is concerned --- the group $G$
is connected. It is a remarkable result --- see,
again,~{\cite{Heyer}} --- the representation kit $\{\bj,\ajk, \eta
\}_{j,k=1}^n$ of a continuous convolution semigroup of measures on
$G$ corresponds to a Gaussian semigroup of measures if and only if
\begin{equation}
\eta=0\ \ \mbox{and}\ \ \big[\ajk\big]_{j,k=1}^n\neq 0.
\end{equation}
This result implies, in particular, that Gaussian semigroups of
measures do exist; precisely, one for each set
$\{\bj,\ajk\}_{j,k=1}^n$, where $[\ajk]_{j,k=1}^n$ is a non-zero
positive matrix. Note, moreover, that the L\'evy-Kintchine
formula~{(\ref{LKF})} holds, in this case, with $\rest=0$, i.e.
\begin{equation} \label{LKFG}
\big(\mathfrak{J}\tre f\big)(g) = \sum_{j=1}^n
\bj\cinque\big(\xi_j\tre f\big)(g) + \sum_{j,k=1}^n
\ajk\cinque\big(\xi_j\tre\xi_k\tre f\big) (g), \ \ \ f\in\cdc(G).
\end{equation}
Note, moreover, that Gaussian semigroups of measures on $G$ form a
special class among the convolution semigroups of measures of the
\emph{first kind} on $G$ (see Sect.~{\ref{basic}).

At this point, in order to get to the main result of this section
(Theorem~{\ref{main-th}} below), we need to pass through four
technical lemmas. We will denote by $\{\mut\}_{t\in \errep}$ an
\emph{arbitrary} continuous convolution semigroup of measures on
$G$, with representation kit $\{\bj,\ajk, \eta \}_{j,k=1}^n$, and by
$\genera$ the infinitesimal generator of the twirling semigroup
associated with the pair $\big(U,\{\mut\}_{t\in \errep}\big)$.
\begin{lemma} \label{lembound}
Let $\varphi\colon G\rightarrow\ccc$ be a bounded Borel function,
which vanishes on a Borel neighborhood of the identity of $G$. Then,
for every sequence $\{\tau_m\}_{m\in\mathbb{N}}$ in $\erreps$
converging to zero, there is a subsequence
$\{\tk\equiv\tau_{m_k}\}_{k\in\mathbb{N}}$ such that the limit
\begin{equation} \label{bou}
\lim_{k\rightarrow\infty}\frac{1}{\tk}\int_G \varphi(g)\; \de\mu_\tk(g)
\end{equation}
exists in $\ccc$.
\end{lemma}

\noindent {\bf Proof:} According to a well known result
--- see~\cite{Heyer}, Lemma~{4.1.4} ---
for every Borel neighborhood of the identity $\mathcal{E}_e$ in $G$,
we have:
\begin{equation}
\sup_{t\in\erreps} t^{-1}
\mut\big(\complement\tre\mathcal{E}_e\big)< \infty .
\end{equation}
Thus, if $\varphi\colon G\rightarrow\ccc$ is a bounded Borel function vanishing on
$\mathcal{E}_e$, we have:
\begin{equation} \label{majo}
\sup_{t\in\erreps} t^{-1} \left|\int_G \varphi(g)\;\de\mut(g)\right|\le
\sup_{t\in\erreps} t^{-1} \int_{\complement\tre\mathcal{E}_e}
\motto|\varphi(g)|\;\de\mut(g)
\le \sup_{g\in G} |\varphi(g)|\ \sup_{t\in\erreps} t^{-1}
\mut\big(\complement\tre\mathcal{E}_e\big) < \infty .
\end{equation}
Now, take any sequence $\{\tau_m\}_{m\in\mathbb{N}}$ in $\erreps$ converging to zero.
Relation~{(\ref{majo})} implies that
\begin{equation} \label{majotau}
\sup_{m\in\mathbb{N}}\tre \frac{1}{\tau_m} \left|\int_G
\varphi(g)\;\de\mu_{\tau_m}(g)\right| < \infty .
\end{equation}
Then, by Bolzano-Weierstrass theorem, there is a subsequence
$\{\tk\equiv\tau_{m_k}\}_{k\in\mathbb{N}}\subset\erreps$ of
$\{\tau_m\}_{m\in\mathbb{N}}$ such that the limit~{(\ref{bou})}
exists in $\ccc$. The proof is complete.~{$\square$}\\
The previous lemma will allow
us to prove the following result,
which will be fundamental for our purposes.

\begin{lemma} \label{biglem}
If $f\colon G\rightarrow\ccc$ is a bounded smooth function such that
the limit
\begin{equation} \label{bound}
\lim_{t\downarrow 0}\frac{1}{t}\Big(\int_G f(g)\; \de\mut(g) - f(e)\Big)
\end{equation}
exists in $\ccc$, then this limit is equal to
\begin{equation} \label{forbiglem}
\sum_{j=1}^{n}\bj\cinque\big(\xi_j\tre f\big)(e) + \mcinque
\sum_{j,k=1}^{n}\ajk\cinque \big(\xi_j\tre\xi_k \tre f\big)(e)  +
\int_\Gst \hspace{-1.3mm} \Big(f(g)-f(e) - \sum_{j=1}^n
\big(\xi_j\tre f\big)(e)\cinque \xe^j(g)\Big) \de\eta(g).
\end{equation}
Therefore, in the case where the convolution semigroup of measures
$\{\mut\}_{t\in \errep}$ is of the first kind (i.e., the associated
L\'evy measure $\eta$ on $\Gst$ is finite), the
limit~{(\ref{bound})} --- if it exists --- is given by
\begin{equation}
\sum_{j=1}^{n}\big(\bj+\cj\big)\tre\big(\xi_j\tre f\big)(e) +
\mcinque \sum_{j,k=1}^{n}\ajk\cinque \big(\xi_j\tre\xi_k \tre
f\big)(e) + \int_\Gst \motto f(g)\; \de\eta(g) - \eta(\Gst)\sei
f(e),
\end{equation}
where:
\begin{equation} \label{coec}
\cj :=-\int_\Gst \motto \xe^j(g)\; \de\eta(g) ,\ \ \ j=1,\ldots, n.
\end{equation}
\end{lemma}

\noindent {\bf Proof:} Since $G$ (being locally compact and second
countable) is $\sigma$-compact, there exists a sequence
$\{\bumpm\}_{m\in\mathbb{N}}$ of non-negative smooth functions on
$G$ characterized as follows:
\begin{enumerate}
\item for every $m\in\mathbb{N}$, $\bumpm$ belongs to
$\cinfc(G;\erre)$ and $\bumpm(G)\subset [0,1]$;

\item there is a sequence $\{\precomp_m\}_{m\in\mathbb{N}}$ of precompact open subsets of $G$ such that
\begin{equation} \label{prec}
e\in\precomp_1,\ \ \ \precomp_1\subset\precomp_2\subset\cdots,\ \ \
\cup_{m=1}^{\infty}\cinque \precomp_m = G,
\end{equation}
\begin{equation} \label{propbump}
\bumpm(g)=1,\ \ \ \forall\quattro g\in\comp_m,
\end{equation}
where $\comp_m$ is the closure of the set $\precomp_m$:
$\comp_m=\overline{\precomp_m}$; we can assume that
\begin{equation} \label{precuno}
\precomp_1\supset\supp\big(\xe^1\big)\cup\ldots\cup\supp\big(\xe^n\big);
\end{equation}

\item there is a sequence $\{\ocomp_m\}_{m\in\mathbb{N}}$ of precompact open subsets of $G$
such that, for every $m\in\mathbb{N}$,
\begin{equation} \label{ocom}
\ocomp_m\supset\comp_m
\end{equation}
and
\begin{equation} \label{propbumpbis}
\bumpm(g)=0,\ \ \ \forall\quattro g\in\complement\tre\ocomp_m .
\end{equation}
\end{enumerate}
In fact, as $G$ is $\sigma$-compact, there exist sequences
$\{\precomp_m\}_{m\in\mathbb{N}}$, $\{\ocomp_m\}_{m\in\mathbb{N}}$
of precompact open subsets of $G$ satisfying~{(\ref{prec})}
and~{(\ref{ocom})}, respectively; relation (\ref{precuno}) can
always be satisfied by the compactness of the supports of the
adapted coordinates. Next, by a standard procedure in the theory of
smooth manifolds one constructs suitable `bump functions'
$\{\bumpm\}_{m\in\mathbb{N}}$, contained in $\cinfc(G;\erre)$,
satisfying~{(\ref{propbump})} and~{(\ref{propbumpbis})}.

By the existence of the limit~{(\ref{bound})}, applying
Lemma~{\ref{lembound}} to the bounded smooth function $f(1-\beta_1)$
(which vanishes on the compact neighborhood $\comp_1$ of $e$), for
some sequence $\{\tk\}_{k\in\mathbb{N}}$ in $\erreps$ converging to
zero we have:
\begin{eqnarray}
\lim_{t\downarrow 0}\frac{1}{t}\Big(\int_G f(g)\; \de\mut(g) -
f(e)\Big) \spa & = & \spa
\lim_{k\rightarrow\infty}\frac{1}{\tk}\Big(\int_G f(g)\;
\de\mu_\tk(g)-
f(e)\Big)\nonumber \\
& = & \spa \lim_{k\rightarrow\infty}\frac{1}{\tk}\Big(\int_G
f(g)\sette \bumpu(g)\; \de\mu_\tk(g)- f(e)\Big)\nonumber \\
\label{giusto}
& + & \spa
\lim_{k\rightarrow\infty}\frac{1}{\tk}\int_G f(g)\sei
(1-\bumpu(g))\; \de\mu_\tk(g),
\end{eqnarray}
where, since the function $f\bumpu$ belongs to $\cinfc(G;\ccc)$ and
$\bumpu(e)=1$, the first limit in the last member
of~{(\ref{giusto})} exists and is equal to
$\big(\mathfrak{J}\quattro (f \bumpu)\big)(e)$, with $\mathfrak{J}$
denoting the generator of the probability semigroup associated with
$\{\mut\}_{t\in\errep}$. We stress that the sequence
$\{\tk\}_{k\in\mathbb{N}}$ can be extracted, as a subsequence, from
any sequence of strictly positive numbers converging to zero. Thus,
we find that
\begin{equation}\label{euno}
\lim_{t\downarrow 0}\frac{1}{t}\Big(\int_G f(g)\; \de\mut(g) -
f(e)\Big) =  \big(\mathfrak{J}\quattro (f \bumpu)\big)(e) +
\lim_{k\rightarrow\infty}\frac{1}{\tk}\int_G f(g)\sei
(1-\bumpu(g))\; \de\mu_\tk(g),
\end{equation}
where, by virtue of the L\'evy-Kintchine formula applied to the
function $f\bumpu\in\cinfc(G;\ccc)$ (note that $(f \bumpu)(g)=f(g)$,
for $g\in\precomp_1$), we can write
\begin{eqnarray}
\big(\mathfrak{J}\quattro (f \bumpu)\big)(e) \spa & = & \spa
\sum_{j=1}^n \bj\cinque\big(\xi_j\tre f\big)(e) + \sum_{j,k=1}^n
\ajk\cinque\big(\xi_j\tre\xi_k\tre f\big) (e)
\nonumber\\
\label{rielabora}
& + & \spa \int_\Gst \hspace{-1.3mm} \Big((f
\bumpu)(g)-f(e) - \sum_{j=1}^n \big(\xi_j\tre f\big)(e)\cinque
\xe^j(g)\Big) \de\eta(g).
\end{eqnarray}

At this point, in order to evaluate the last term
in~{(\ref{rielabora})}, it will be convenient to set
\begin{eqnarray}
\varphi_{1,1}(g) \spa & \equiv & \spa f(g)\sette
\bumpu(g),\hspace{1.2mm} \mbox{and, for $m\in\mathbb{N}$, $m\ge 2$,}
\\ \nonumber
\varphi_{1,1}(g) \spa & = & \spa \varphi_{m,1}(g) +
\varphi_{m,2}(g),\ \varphi_{m,1}(g):= f(g) \sette \bumpm(g),\
\varphi_{m,2}(g):= f(g)\sei (\bumpu (g)-\bumpm(g)).
\end{eqnarray}
Clearly, the functions $\{\varphi_{m,1}\}_{m\ge 1}$ belong to $\cinfc(G;\ccc)$.
It is easy to check that the functions
$\{\varphi_{m,2}\}_{m\ge 2}$ belong to $\cinfc(G;\ccc)$, as well. Indeed, they are
obviously smooth and
\begin{equation}
\supp(\bumpu -
\bumpm)\subset\overline{\complement\tre\comp_1\cap\big(\ocomp_1\cup\ocomp_{m}\big)}
\subset \overline{\complement\tre\comp_1}\cap
\overline{\big(\ocomp_1\cup\ocomp_{m}\big)}=\overline{\complement\tre\comp_1}\cap
\big(\overline{\ocomp_1}\cup\overline{\ocomp_{m}}\big).
\end{equation}
Thus, the set $\supp(\bumpu - \bumpm)$ is compact in $G$. Note that,
as it does not contain the identity, it is a compact set in $\Gst$,
as well; hence: $\{\varphi_{m,2}\}_{m\ge
2}\subset\cinfc(\Gst;\ccc)$. This fact allows us to use
formula~{(\ref{caralevi})} in such a way to decompose the last term
in~{(\ref{rielabora})} as follows:
\begin{eqnarray}
\int_\Gst \hspace{-1.3mm} \Big(\varphi_{1,1}(g)-f(e) - \sum_{j=1}^n
\big(\xi_j\tre f\big)(e)\cinque \xe^j(g)\Big) \de\eta(g) \spa & = &
\spa \int_\Gst \hspace{-1.3mm} \Big(\varphi_{m,1}(g)-f(e) -
\sum_{j=1}^n \big(\xi_j\tre f\big)(e)\cinque \xe^j(g)\Big)
\de\eta(g)
\nonumber \\ \label{edue}
& + & \spa \lim_{k\rightarrow\infty}\frac{1}{\tk}\int_G
\varphi_{m,2}(g) \; \de\mu_\tk(g)\equiv\varkappa,\ \ \ m\ge 2.
\end{eqnarray}
Note that the number $\varkappa$ does not depend on the index $m$.
At this point, considering the last term in~{(\ref{euno})}, for every
$m\ge 2$ we have:
\begin{eqnarray}
\varkappa + \lim_{k\rightarrow\infty}\frac{1}{\tk}\int_G f(g)\sei
(1-\bumpu(g))\; \de\mu_\tk(g) \spa & = & \spa  \int_\Gst
\hspace{-1.3mm} \Big(\varphi_{m,1}(g)-f(e) - \sum_{j=1}^n
\big(\xi_j\tre f\big)(e)\cinque \xe^j(g)\Big) \de\eta(g) \nonumber
\\ \label{etre} & + & \spa
\lim_{k\rightarrow\infty}\frac{1}{\tk}\int_G f(g)\sei(1-\bumpm(g))
\; \de\mu_\tk(g).
\end{eqnarray}

The r.h.s.\ of relation~{(\ref{etre})} can be regarded as the (constant)
sum of two sequences labeled by the index $m$. Therefore, if one of the two sequences
is converging, the other one must converge too.
Let us prove that the limit
\begin{equation} \label{exilim}
\lim_{m\rightarrow\infty} \int_\Gst \hspace{-1.3mm}
\Big(\varphi_{m,1}(g)-f(e) - \sum_{j=1}^n \big(\xi_j\tre
f\big)(e)\cinque \xe^j(g)\Big) \de\eta(g)
\end{equation}
exists and is equal to
\begin{equation} \label{equilim}
\int_\Gst
\hspace{-1.3mm} \Big(f(g)-f(e) - \sum_{j=1}^n \big(\xi_j\tre
f\big)(e)\cinque \xe^j(g)\Big) \de\eta(g).
\end{equation}
Indeed --- observing that, by~{(\ref{precuno})},
$\xe^j(g)=\xe^j(g)\sette \bumpm(g)$, and denoting by
$\chi_{\complement\tre\comp_{m}}\hspace{-1mm}$ the characteristic
function of the set $\complement\tre\comp_{m}$ --- we can write the
estimate
\begin{eqnarray} \hspace{-10.5mm}
\Big|\varphi_{m,1}(g) - f(e) - \sum_{j=1}^n \big(\xi_j\tre
f\big)(e)\cinque \xe^j(g)\Big| \spa & = & \spa \Big|f(g)\sette
\bumpm(g)-f(e) - \sum_{j=1}^n \big(\xi_j\tre f\big)(e)\cinque
\xe^j(g)\sette \bumpm(g)\Big|
\nonumber \\
& \le & \spa \Big|f(g)-f(e) - \sum_{j=1}^n \big(\xi_j\tre
f\big)(e)\cinque \xe^j(g)\Big|\sette \bumpm(g)
\nonumber \\
& + & \spa |f(e)|\sette (1- \bumpm(g)) \nonumber \\
& \le & \spa \Big|f(g)-f(e) - \sum_{j=1}^n \big(\xi_j\tre
f\big)(e)\cinque \xe^j(g)\Big| +
|f(e)|\nove\chi_{\complement\tre\comp_{m}}\msette(g),
\end{eqnarray}
for all $m\in\mathbb{N}$ and $g\in G$. Therefore, since
$\chi_{\complement\tre\comp_{m}}\hspace{-1mm}\le
\chi_{\complement\tre\comp_{1}}\hspace{-1mm}$, we find out that
\begin{equation} \label{lastlast}
\Big|\varphi_{m,1}(g) - f(e) - \sum_{j=1}^n \big(\xi_j\tre
f\big)(e)\cinque \xe^j(g)\Big| \le \Big|f(g)-f(e) - \sum_{j=1}^n
\big(\xi_j\tre f\big)(e)\cinque \xe^j(g)\Big| +
|f(e)|\nove\chi_{\complement\tre\comp_{1}}\msette(g).
\end{equation}
The expression on the r.h.s.\ of~{(\ref{lastlast})} defines a
function contained in $\mathrm{L}^1(\Gst,\eta;\ccc)$. Therefore,
since $\lim_{m\rightarrow\infty}\bumpm (g) = 1$, for all $g\in G$,
by the `dominated convergence theorem' the limit~{(\ref{exilim})}
exists and is equal to~{(\ref{equilim})}, as claimed.

Let us resume what we have obtained up to this point. By
relations~{(\ref{euno})}, (\ref{rielabora}), (\ref{edue})
and~{(\ref{etre})}, and by the fact that the limit~{(\ref{exilim})}
is equal to~{(\ref{equilim})}, we conclude that the existence of the
limit~{(\ref{bound})}, for a bounded smooth function $f\colon
G\rightarrow\ccc$, implies that this limit must coincide with
\begin{eqnarray} \label{somma}
\sum_{j=1}^{n}\bj\cinque\big(\xi_j\tre f\big)(e) + \mcinque
\sum_{j,k=1}^{n}\ajk\cinque \big(\xi_j\tre\xi_k \tre f\big)(e) \spa
& + & \spa \int_\Gst \hspace{-1.3mm} \Big(f(g)-f(e) - \sum_{j=1}^n
\big(\xi_j\tre f\big)(e)\cinque \xe^j(g)\Big) \de\eta(g)
\nonumber\\
& + & \spa \lim_{m\rightarrow\infty}
\lim_{k\rightarrow\infty}\frac{1}{\tk}\int_G f(g)\sei(1-\bumpm(g))
\; \de\mu_\tk(g),
\end{eqnarray}
for some sequence $\{\tk\}_{k\in\mathbb{N}}$ in $\erreps$ converging
to zero that can be extracted, as a subsequence, from any sequence
of strictly positive numbers converging to zero. Note that the
iterated limit above \emph{must} exist (as the first member
of~{(\ref{etre})} does not depend on $m$ and the
limit~{(\ref{exilim})} exists).

We now apply this result to the function $f\equiv 1$. Then, we find
immediately that
\begin{equation} \label{nova}
\lim_{m\rightarrow\infty}\lim_{k\rightarrow\infty}\tre\frac{1}{\tau_k}\int_G
(1-\bumpm(g)) \; \de\mu_{\tau_k}(g)=0 ,
\end{equation}
for some sequence $\{\tau_k\}_{k\in\mathbb{N}}$ in $\erreps$
converging to zero.

Finally, considering again an \emph{arbitrary} bounded smooth
function $f$ on $G$ for which the limit~{(\ref{bound})} exists,
extract from $\{\tau_k\}_{k\in\mathbb{N}}$ a subsequence
$\{\tk\}_{k\in\mathbb{N}}$ such that this limit coincides
with~{(\ref{somma})}. From~{(\ref{nova})}
--- observing that the inequality
\begin{equation}
\left| \int_G f(g)\sei(1-\bumpm(g)) \; \de\mu_\tk(g) \right|\le
\|f\supn \int_G (1-\bumpm(g)) \; \de\mu_\tk(g)
\end{equation}
implies
\begin{eqnarray} \hspace{-9mm}
\left| \lim_{m\rightarrow\infty}
\lim_{k\rightarrow\infty}\frac{1}{\tk}\int_G f(g)\sei(1-\bumpm(g))
\; \de\mu_\tk(g)\right| \spa & = & \spa \lim_{m\rightarrow\infty}\lim_{k\rightarrow\infty}\frac{1}{\tk}
\left|\int_G f(g)\sei(1-\bumpm(g))
\; \de\mu_\tk(g)\right|
\nonumber\\
& \le & \spa \|f\supn \lim_{m\rightarrow\infty}
\lim_{k\rightarrow\infty}\frac{1}{\tk} \int_G (1-\bumpm(g)) \; \de\mu_\tk(g)
\end{eqnarray}
--- we conclude that the last term in~{(\ref{somma})} vanishes and the proof is
complete.~{$\square$}\\
The next lemma will lead us very close to the main result of this
section.

\begin{lemma} \label{veryclose}
With the previous notations and assumptions, for every operator
$\opa\in\bH$, the following relation holds:
\begin{eqnarray} \hspace{-9mm}
\genera\tre\opa \spa & = & \spa \lim_{t\downarrow
0}\frac{1}{t}\Big(\int_G \de\mut(g)\ U(g)\tre \opa\sei U(g)^\ast
-\opa\Big)  \nonumber\\ & = & \spa
\sum_{j=1}^{n}\bj\cinque[\opx_j,\opa\big] + \mcinque
\sum_{j,k=1}^{n}\ajk\cinque
\big(\big\{\opx_j\tre\opx_k,\opa\big\}-2\tre\opx_j\tre\opa\sei\opx_k\big)
\nonumber \\ \label{gentot} & + & \spa \int_\Gst \hspace{-1.3mm}
\Big( U(g)\tre \opa\sei U(g)^\ast -\opa - \sum_{j=1}^n
\xe^j(g)\cinque\big[\opx_j,\opa\big]\Big)
\de\eta(g)\equiv\opa^\prime,
\end{eqnarray}
where $\{\cdot,\cdot\}$ is the anti-commutator and the set
$\{\opx_1,\ldots,\opx_n\}\subset\ah$ is the $n$-tuple of operators
defined by~{(\ref{opxs})}. Suppose, in particular, that
$\{\mut\}_{t\in \errep}$ is a convolution semigroup of measures of
the first kind. Then, for every $\opa\in\bH$, we have:
\begin{eqnarray}
\lim_{t\downarrow 0}\frac{1}{t}\Big(\int_G \de\mut(g)\ U(g)\tre
\opa\sei U(g)^\ast -\opa\Big) \hspace{-0.8mm}  \spa & = & \spa
\sum_{j=1}^{n}\big(\bj+\cj\big)\tre\big[\opx_j,\opa\big] + \mcinque
\sum_{j,k=1}^{n}\ajk\cinque
\big(\big\{\opx_j\tre\opx_k,\opa\big\}-2\tre\opx_j\tre\opa\sei\opx_k\big)
\nonumber\\ \label{gentot-bis} & + & \spa
\eta(\Gst)\tre\big(\randue-I\big)\tre \opa ,
\end{eqnarray}
where $\{\cj\}_{j=1}^n$ are real numbers defined by~{(\ref{coec})},
and $\randue\colon\bH\rightarrow\bH$ is identically zero for
$\eta=0$ and a random unitary map for $\eta\neq 0$, with
\begin{equation}
\randue = \eta(\Gst)^{-1}\int_\Gst \motto \de\eta(g)\ \urep(g),\ \ \
\eta\neq 0.
\end{equation}
\end{lemma}

\noindent {\bf Proof:} It is sufficient to show that
\begin{equation} \label{prova}
\big\langle \phi,\big(\genera\tre\opa\big)\tre \psi\big\rangle =
\lim_{t\downarrow 0}\frac{1}{t}\Big(\int_G \de\mut(g)\ \langle \phi,
U(g)\tre \opa\sei U(g)^\ast \psi\rangle -\langle \phi,\opa\cinque
\psi\rangle\Big)\hspace{-0.8mm} =  \langle \phi,\opa^\prime\tre
\psi\rangle,
\end{equation}
for arbitrary $\opa\in\bH$ and $\phi,\psi\in\hh$, where
$\opa^\prime$ is the shorthand notation introduced
in~{(\ref{gentot})}. To this aim, since the limit in~{(\ref{prova})}
exists, we can apply Lemma~{\ref{biglem}} to the bounded smooth
function $f\colon G\rightarrow\ccc$ defined by
\begin{equation}
f(g):=\langle \phi, U(g)\tre \opa\sei U(g)^\ast \psi\rangle.
\end{equation}
Using the notation introduced in Sect.~{\ref{basic}}, there exists a
neighborhood of the identity $\mathcal{E}_e$ in $G$ such that
\begin{equation}
U(g)=\eee^{\xe^1(g) \cinque \opx_1 + \cdots + \xe^n(g) \cinque
\opx_n},\ \ \ \forall\quattro g\in\mathcal{E}_e.
\end{equation}
Therefore, we have that
\begin{equation}
\xi_j\sei \langle \phi, U(g)\tre \opa\sei U(g)^\ast\tre
\psi\rangle\quattro\Big\vert_{g=e} =  \big\langle \phi,
\big[\opx_j,\opa\big]\tre\psi\big\rangle,
\end{equation}
\begin{equation}
\xi_j\tre\xi_k\sei \langle \phi, U(g)\tre \opa\sei U(g)^\ast\tre
\psi\rangle \quattro\Big\vert_{g=e} =  \big\langle
\phi,\big(\opx_j \tre\opx_k \tre\opa + \opa\sei\opx_k\tre\opx_j-\opx_j\tre\opa\sei\opx_k -
\opx_k\tre\opa\sei\opx_j \big)\tre\psi\big\rangle.
\end{equation}
Now, exploiting formula~{(\ref{forbiglem})} and the fact that the
matrix $\left[\ajk\right]_{\hspace{-0.5mm}j,k=1}^n$ is symmetric, we
obtain immediately relation~{(\ref{prova})}.~{$\square$}\\
The last technical lemma will establish a useful link between the
generator of the twirling semigroup associated with the pair
$\big(U,\{\mut\}_{t\in \errep}\big)$ --- with $\{\mut\}_{t\in
\errep}$ denoting a \emph{generic} continuous convolution semigroup
of measures on $G$ --- and the convolution semigroups of measures on
$G$ \emph{of the first kind}.

\begin{lemma} \label{limeas}
There exists a sequence $\big\{\{\mu_{t;\tre m}\}_{t\in\errep}\colon
m\in\mathbb{N}\big\}$ of continuous convolution semigroups of
measures of the first kind on $G$ --- with $\{\mu_{t;\tre
m}\}_{t\in\errep}$ having a representation kit of the form
$\{\bj,\ajk, \eta_m \}_{j,k=1}^n$ --- such that
\begin{equation}
\lim_{m\rightarrow\infty}\generam=\genera ,\ \mbox{and}\ \;
\lim_{m\rightarrow\infty}\int_\Gst \motto f(g)\; \de\eta_m(g)
=\int_\Gst \motto f(g)\; \de\eta(g),
\end{equation}
for every bounded Borel function $f\colon \Gst \rightarrow\ccc$
belonging to $\mathrm{L}^1(\Gst,\eta;\ccc)$.
\end{lemma}

\noindent {\bf Proof:} Let $\{\bj,\ajk, \eta \}_{j,k=1}^n$ denote,
as usual, the representation kit of the convolution semigroup of
measures $\{\mut\}_{t\in\errep}$, and let $\Phi\colon
G\rightarrow\errep$ be a Hunt function and $\Phi^\prime$ its
restriction to $\Gst$. For every $m\in\mathbb{N}$, consider the
measure $\eta_m$ on $\Gst$ determined by
\begin{equation}
\de\eta_m(g)=\big(1-\exp\big(\msette-m\quattro\Phi^\prime(g)\big)\big)\quattro\de\eta(g),\
\ \ g\in\Gst.
\end{equation}
The measure $\eta_m$ is finite (by construction), for all
$m\in\mathbb{N}$, and, as
$0\le\big(1-\exp\big(\msette-m\quattro\Phi^\prime(g)\big)\big)\le
1$, by the `dominated convergence theorem' we have that
\begin{equation} \label{pseudoweak}
\lim_{m\rightarrow\infty}\int_\Gst \motto f(g)\; \de\eta_m(g)
=\int_\Gst \motto f(g)\; \de\eta(g),
\end{equation}
for every bounded Borel function $f\colon \Gst \rightarrow\ccc$
contained in $\mathrm{L}^1(\Gst,\eta;\ccc)$. Denote by
$\{\mu_{t;\tre m}\}_{t\in\errep}$ the continuous convolution
semigroup of measures with representation kit $\{\bj,\ajk, \eta_m
\}_{j,k=1}^n$. From relations~{(\ref{gentot})}
and~{(\ref{pseudoweak})} --- setting $f(g)=\big\langle\phi, U(g)\tre
\opa\sei U(g)^\ast -\opa - \sum_{j=1}^n
\xe^j(g)\cinque\big[\opx_j,\opa\big]\psi\big\rangle$, $g\in \Gst$,
for any $\opa\in\bH$ and $\phi,\psi\in\hh$ ---  we obtain that
\begin{equation}
\lim_{m\rightarrow\infty}\generam=\genera.
\end{equation}
The proof is complete.~{$\square$}

Having completed the main technical proofs, we are finally ready to
focus on the main result of this section, which can be regarded as a
generalization of an already cited classical result of K\"ummerer
and Maassen~{\cite{Kummerer}}. The latter result is obtained from
the former (namely, Theorem~{\ref{main-th}} below) by choosing the
unitary representation $U$ as the defining representation of
$\sunig(\dime)$ (up to unitary equivalence). It will be now
convenient to establish a few additional notations. Given a nonempty
subset $\subs$ of $\uni(\hh)$, we will denote by $\cone(\subs)$ the
cone in $\supops$ generated by this set --- i.e.,
$\cone(\subs):=\errep\tre\subs$ --- and by $\clcone(\subs)$ the
closure of such cone. If $0\in\subs$, consider, moreover, the set
\begin{equation} \label{defzercone}
\zercone (\subs) :=
\big\{\mathfrak{A}\in\supops\colon\exists\tre\{\alpha_m\}_{m\in\mathbb{N}}\subset\erreps,\
\alpha_m\rightarrow \infty,\
\exists\tre\{\mathfrak{A}_m\}_{m\in\mathbb{N}}\subset\subs
\hspace{2.2mm} \mbox{s.t.}\
\alpha_m\tre\mathfrak{A}_m\rightarrow\mathfrak{A}\big\}.
\end{equation}
It can be shown that if $\subs$ is a closed set, then $\zercone
(\subs)$ is a closed cone (see~{\cite{Durea}}, where a closed subset
of a normed vector space is considered). Denoting, as above,
by $\suni$ a subgroup of $\uni(\hh)$ and by
$\csuni$ the subgroup of $\uni(\hh)$ which is the closure of
$\suni$, the sets $\clcone(\cvvv -I)$ and
$\coco(\suni):=\clcocone(\vvv -I)$ are characterized as follows.
\begin{proposition}
For the closed convex cone $\coco(\suni):=\clcocone(\vvv -I)$ we
have:
\begin{equation} \label{propcoco}
\coco(\suni)= \clcocone(\cvvv -I)=\clcocone(\dmru(\suni) -I) =
\clcocone(\dmru(\csuni) -I).
\end{equation}
The set $\zercone(\cvvv -I)$ is a closed cone in $\supops$.
The closed cone $\clcone(\cvvv -I)$ is contained in $\coco(\suni)$ and
\begin{equation} \label{unicone}
\clcone(\cvvv -I) = \zercone(\cvvv
-I)\cup \tre\cone(\cvvv -I).
\end{equation}
\end{proposition}

\noindent {\bf Proof:}
The proof of relations~{(\ref{propcoco})} goes as follows. First
observe that
\begin{equation}
\coco(\suni):=\clcocone(\vvv -I)=\clcocone(\overline{\vvv
-I})=\clcocone(\cvvv -I).
\end{equation}
Next, we have:
\begin{eqnarray}
\clcocone(\vvv -I) \spa & = & \spa \clcocone(\co(\vvv -I)) \\
\nonumber & = & \spa \clcocone(\dmru(\suni) -I) =
\clcocone(\overline{\dmru(\suni) -I})= \clcocone(\dmru(\csuni) -I).
\end{eqnarray}
Thus, the proof of~{(\ref{propcoco})} is complete.

Next, since $\cvvv -I$ is a closed set, $\zercone(\cvvv -I)$ is a
closed cone, and from our previous arguments it is clear that the
closed cone $\clcone(\cvvv -I)$ is contained in $\coco(\suni)$. Let
us prove relation~{(\ref{unicone})}. For every compact subset
$\mathcal{K}$ of $\supops$ such that $0\in\mathcal{K}$, the
following decomposition holds:
$\clcone(\mathcal{K})=\zercone(\mathcal{K})+\cone(\mathcal{K})$
(see~{\cite{Durea}}, Theorem~{3.2}, and take into account the fact
that the `asymptotic cone' --- or `recession cone' --- generated by
a bounded set coincides with the origin). Apply this result to the
compact set $\cvvv -I$. The proof is complete.~{$\square$}\\
In the following, the subgroups $\suni$ and $\csuni$ of $\uni(\hh)$
will be identified with the subgroups $\ug$ and $\cug$,
respectively. Let $\vu$ be the real vector space obtained by
projecting $\ima\big(\ran(\pu)\big)$
--- regarded as a vector subspace of $\brH$ --- onto the orthogonal
complement of the one-dimensional space spanned by the identity;
namely,
\begin{equation}
\vu:=\big\{\opa\in\brH\colon\ \opa
=\ima\big(\pu(\xi)-\dime^{-1}\tr(\pu(\xi))I\big),\ \xi\in\lie\big\}.
\end{equation}
We will denote by $\dimu$ the dimension of the vector space $\vu$
($\dimu\le\min \{n,\dime-1\}$). Observe that, if $G$ is a semisimple
Lie group, then $[\lie,\lie]=\lie$ and
$\vu=\ima\big(\ran(\pu)\big)$. Finally, in the case where
$\{\mut\}_{t\in \errep}$ is of regular type, the adapted coordinates
$\{g\mapsto \xe^1(g), \ldots, g\mapsto \xe^n(g)\}$ are integrable
with respect to the L\'evy measure $\eta$ and we can set
\begin{equation}
\cj :=-\int_\Gst \motto \xe^j(g)\; \de\eta(g) ,\ \ \ j=1,\ldots, n.
\end{equation}

\begin{theorem} \label{main-th}
Let $G$ be a Lie group and $U$ a smooth unitary representation of
$G$ in the Hilbert space $\hh$.  Then, for every continuous
semigroup of measures $\{\mut\}_{t\in \errep}$ on $G$ --- let
$\{\bj,\ajk, \eta \}_{j,k=1}^n$ be the associated representation kit
--- the infinitesimal generator $\genera\colon\bH\rightarrow\bH$ of
the twirling semigroup $\{\twis\}_{t\in\errep}$ associated with the
pair $(\{\mut\}_{t\in\errep}, U)$ is of the form
\begin{equation} \label{decompo}
\genera =\generag + \generaw,
\end{equation}
where $\generag$ and $\generaw$ belong to the closed convex cone
$\coco(\ug)\subset\coconeh\subset\supops$ and are given by
\begin{equation} \label{defigenerag}
\generag := \sum_{j=1}^{n}\bj\cinque\big[\opx_j,(\cdot)\big] +
\mcinque \sum_{j,k=1}^{n}\ajk\cinque
\big(\big\{\opx_j\tre\opx_k,(\cdot)\big\}-2\tre\opx_j\tre(\cdot)\sei\opx_k\big),
\end{equation}
\begin{equation}
\generaw := \int_\Gst \hspace{-1.3mm} \Big( \urep(g) - I -
\sum_{j=1}^n \xe^j(g)\cinque\big[\opx_j,(\cdot)\big]\Big)
\de\eta(g),
\end{equation}
with $\opx_1,\ldots,\opx_n$ the skewadjoint operators defined
by~{(\ref{opxs})}. In the case where the semigroup of measures
$\{\mut\}_{t\in \errep}$ is of the first kind, we have:
\begin{equation} \label{partic}
\generaw = \eta(\Gst)\tre\big(\randue-I\big)
+\sum_{j=1}^{n}\cj\cinque\big[\opx_j,(\cdot)\big],
\end{equation}
with $\randue\colon\bH\rightarrow\bH$ identically zero, for
$\eta=0$, and
\begin{equation}
\randue := \eta(\Gst)^{-1}\int_\Gst \motto\urep(g)\;
\de\eta(g)\in\dmru(\cug),\ \ \mbox{for $\eta\neq 0$}.
\end{equation}
Suppose, instead, that the semigroup of measures $\{\mut\}_{t\in
\errep}$ is of the second kind. Then, there exists a sequence
$\big\{\{\mu_{t;\tre m}\}_{t\in\errep}\colon m\in\mathbb{N}\big\}$
of continuous convolution semigroups of measures of the first kind
on $G$ --- with $\{\mu_{t;\tre m}\}_{t\in\errep}$ having a
representation kit of the form $\{\bj,\ajk, \eta_m \}_{j,k=1}^n$ ---
such that $\lim_{m\rightarrow\infty}\generam=\genera$, and
$\lim_{m\rightarrow\infty}\int_\Gst \motto f(g)\; \de\eta_m(g)
=\int_\Gst \motto f(g)\; \de\eta(g)$, for every bounded Borel
function $f\colon \Gst \rightarrow\ccc$ belonging to
$\mathrm{L}^1(\Gst,\eta;\ccc)$. Moreover, we have that
\begin{equation} \label{sorto}
\generaw =
\lim_{m\rightarrow\infty}\Big(\eta_m(\Gst)\tre\big(\randuem-I\big) +
\sum_{j=1}^{n}\cjm\cinque\big[\opx_j,(\cdot)\big]\Big),
\end{equation}
and, in the case where
$\{\mut\}_{t\in \errep}$ is of regular type,
\begin{equation} \label{sorta}
\generaw = \generawo +
\sum_{j=1}^{n}\cj\cinque\big[\opx_j,(\cdot)\big],
\end{equation}
with $\generawo$ denoting the element of the closed convex cone
$\coco(\ug)$ determined by
\begin{equation} \label{sortb}
\generawo =
\lim_{m\rightarrow\infty}\eta_m(\Gst)\tre\big(\randuem-I\big).
\end{equation}
The superoperator defined by~{(\ref{defigenerag})} can be expressed
in the canonical form
\begin{equation} \label{pseudogauss}
\generag = - \ima \big[\oph, (\cdot)\big] + \sum_{k=1}^{\dimu}
\gamma_k\Big(\opf_k^{\phantom{\ast}}(\cdot)\sei\opf_{k}^{\phantom{\ast}}-\frac{1}{2}
\big(\opf_k^2\tre(\cdot)+ (\cdot)\sei\opf_k^2\big)\Big),\ \ \
\gamma_k\ge 0,
\end{equation}
with $\oph, \opf_1 \fast,\ldots,\opf_\dimu \fast$ traceless
selfadjoint operators in $\hh$ satisfying
\begin{equation}
\oph, \opf_1\fast,\ldots,\opf_\dimu\fast\subset \vu,\ \ \
\big\langle\opf_{j}^{\phantom{\ast}},\opf_k^{\phantom{\ast}}\big\ranglehs
=\delta_{jk},\ \  j,k =1,\dots,\dimu.
\end{equation}
In particular, if $\{\mut\}_{t\in \errep}$ is a Gaussian semigroup
of measures, then  $\generaw=0$ and $\generag$ is of the
form~{(\ref{forgeneg})}, i.e., $\{\twis\}_{t\in\errep}$ is a
Gaussian dynamical semigroup. Finally, for every superoperator
$\gene\colon\bH\rightarrow\bH$ of the form $\gene=\geneg +
\gamma_0\big(\randu -I\big)$ --- with $\geneg$ of the general form
given by the r.h.s.\ of~{(\ref{pseudogauss})}, $\randu$ belonging to
$\dmru(\ug)$ and $\gamma_0\ge 0$ --- there is a continuous
convolution semigroup of measures $\{\mut\}_{t\in \errep}$ on $G$
--- with associated L\'evy measure identically zero, if $\gamma_0\big(\randu -I\big)=0$
--- such that the infinitesimal generator of the twirling semigroup
$\{\twis\}_{t\in\errep}$ induced by the pair
$(\{\mut\}_{t\in\errep}, U)$ is $\gene$.
\end{theorem}

\noindent {\bf Proof:} By Lemma~{\ref{veryclose}}, the infinitesimal
generator $\genera$ is of the form~{(\ref{decompo})}. In particular,
in the case where the semigroup of measures $\{\mut\}_{t\in \errep}$
is of the first kind, the superoperator $\generaw$ is of the
form~{(\ref{partic})}. By Lemma~{\ref{limeas}}, in the case where
the semigroup of measures $\{\mut\}_{t\in \errep}$ is of the second
kind, there exists a sequence $\big\{\{\mu_{t;\tre
m}\}_{t\in\errep}\colon m\in\mathbb{N}\big\}$ of continuous
convolution semigroups of measures of the first kind on $G$ --- with
$\{\mu_{t;\tre m}\}_{t\in\errep}$ having a representation kit of the
form $\{\bj,\ajk, \eta_m \}_{j,k=1}^n$ --- such that
\begin{equation}
\lim_{m\rightarrow\infty}\generam=\genera ,\ \mbox{and}\
\lim_{m\rightarrow\infty}\int_\Gst \motto f(g)\; \de\eta_m(g)
=\int_\Gst \motto f(g)\; \de\eta(g),
\end{equation}
for every bounded Borel function $f\colon \Gst \rightarrow\ccc$
belonging to $\mathrm{L}^1(\Gst,\eta;\ccc)$. It follows
that~{(\ref{sorto})} --- and, in the case where $\{\mut\}_{t\in
\errep}$ is of regular type, as $\lim_{m\rightarrow\infty}\cjm=\cj$,
(\ref{sorta}) --- hold true.

Let us prove that the superoperators $\generag$ and $\generaw$ of
decomposition~{(\ref{decompo})} belong to the convex cone
$\coco(\ug)$. Indeed, diagonalizing the positive matrix
$\left[\ajk\right]_{\hspace{-0.5mm}j,k=1}^n$ and introducing a
suitable new basis $\{\upsilon_1,\ldots,\upsilon_n\}$ in $\lie$, we
can write $\generag$ in the form
\begin{equation}
\generag := \big[\opy_0,(\cdot)\big] + \mcinque \sum_{j=1}^{n}
\lambda_j
\big(\big\{\opy_j\tre\opy_j,(\cdot)\big\}-2\tre\opy_j\tre(\cdot)\sei\opy_j\big),\
\ \ \ \lambda_j\ge 0,
\end{equation}
where $\opy_0\in\ran(\pu)$, $\opy_0=\pu(\upsilon_0)$ (for some
$\upsilon_0\in\lie$), and
$\opy_1=\pu(\upsilon_1),\ldots,\opy_n=\pu(\upsilon_n)$ are
skewadjoint operators in $\hh$. For the superoperator
$\big[\opy_0,(\cdot)\big]$ we have:
\begin{equation}
\big[\opy_0,(\cdot)\big]=\frac{\de}{\de
t}\tre\big(\eee^{t\opy_0}(\cdot)\sette\eee^{-t\opy_0}\big)\Big\vert_{t=0}
=\lim_{t\downarrow
0}t^{-1}\big(\eee^{t\opy_0}(\cdot)\sette\eee^{-t\opy_0}
-(\cdot)\big),\ \ \ \eee^{\pm t\opy_0}=U(\exp_G(\pm t\upsilon_0)).
\end{equation}
Therefore, $[\opy_0,(\cdot)\big]$ belongs to $\coco(\ug)$.
Analogously, since $\eee^{\pm t\opy_j}=U(\exp_G(\pm t\upsilon_j))$,
we have that
\begin{eqnarray} \hspace{-8mm}
\big\{\opy_j\tre\opy_j,(\cdot)\big\}-2\tre\opy_j\tre(\cdot)\sei\opy_j
\spa & = & \spa \frac{1}{2} \tre
\big[\opy_j,\big[\opy_j,(\cdot)\big]\big]
\nonumber\\
& = & \spa \frac{1}{2}\tre \frac{\de^2}{\de t^2}\tre
\big(\eee^{t\opy_j}(\cdot)\sette\eee^{-t\opy_j}\big)\Big\vert_{t=0}
\nonumber\\
& = & \spa \lim_{t\downarrow
0}\frac{1}{2t^2}\Big(\big(\eee^{t\opy_j}(\cdot)\sette\eee^{-t\opy_j}
-(\cdot)\big) + \big(\eee^{-t\opy_j}(\cdot)\sette\eee^{t\opy_j}
-(\cdot)\big)\Big)\in\coco(\ug) .
\end{eqnarray}
Hence, $\generag$ is a convex combination of elements of the closed
convex cone $\coco(\ug)$. By a similar argument $\generaw$ belongs
to $\coco(\ug)$, as well.

The canonical form~{(\ref{pseudogauss})} of the superoperator
$\generag$ follows from a direct calculation (hint: expand the
selfadjoint operators $\ima\opx_1,\ldots,\ima\opx_n$ with respect to
an orthonormal basis in $\brH$ including a multiple of the identity,
and exploit the fact that $\big[\ajk\big]_{j,k=1}^n$ is a positive
symmetric matrix). If $\{\mut\}_{t\in \errep}$ is a Gaussian
semigroup of measures, then the associated L\'evy measure is
identically zero and $\big[\ajk\big]_{j,k=1}^n\neq 0$. Therefore, in
this case, $\generaw=0$ and $\generag$ must be of the
form~{(\ref{forgeneg})}.

Let us prove the last assertion of the theorem. First, if $\gamma_0\big(\randu -I\big)\neq 0$,
choose a L\'evy measure
$\eta$ (of the first kind) on $\Gst$ as a superposition of point mass measures in such a way that
\begin{equation}
\int_\Gst \hspace{-1.3mm} \Big( \urep(g) - I\Big) \de\eta(g) = \gamma_0\big(\randu -I\big),\ \ \
(\eta(\Gst)=\gamma_0);
\end{equation}
otherwise set $\eta =0$. Next, take vectors $\zeta_0,\zeta_1,\ldots,\zeta_\dimu$ in $\lie$
such that
\begin{equation}
\zeta_0\in\Big(\pu^{-1}\big(\hspace{-0.9mm}
-\ima\tre\projinv\big(\oph \big)\big)- \sum_{j=1}^n\tre \cj\tre\xi_j \Big),\ \ \
\zeta_k\in\pu^{-1}\big(\ima\tre\projinv\big(\opf_k\big)\big),\ k=1,\ldots,\dimu,
\end{equation}
where $\proju$ is the orthogonal projection (with respect to the Hilbert-Schmidt scalar product)
of $\brH$ onto $\vu$. Now, expand the vectors $\zeta_0,\zeta_1,\ldots,\zeta_\dimu$ with respect to
the basis $\{\xi_1,\ldots,\xi_n\}$ in $\lie$: $\zeta_0=\sum_{j=1}^n b^j\tre\xi_j$,
$\zeta_k=\sum_{l=1}^n d_{kl}\tre\xi_l$, $k=1,\ldots,\dimu$. At this point, one can check that
\begin{eqnarray}
\gene=\geneg +
\gamma_0\big(\randu -I\big)  \spa & = & \spa
\sum_{j=1}^{n}\bj\cinque\big[\opx_j,(\cdot)\big] +
\mcinque \sum_{j,k=1}^{n}\ajk\cinque
\big(\big\{\opx_j\tre\opx_k,(\cdot)\big\}-2\tre\opx_j\tre(\cdot)\sei\opx_k\big)
\nonumber\\
& + & \spa \label{compa}
\int_\Gst \hspace{-1.3mm} \Big( \urep(g) - I\Big) \de\eta(g)
+\sum_{j=1}^{n}\cj\cinque\big[\opx_j,(\cdot)\big],
\end{eqnarray}
where $\big[\ajk\big]_{j,k=1}^n$ is the positive real matrix defined by
\begin{equation}
\ajk := \frac{1}{2}\sum_{l,m=1}^{\dimu} \gamma_l\sei \delta_{lm}\sei d_{lj}\sei d_{mk}.
\end{equation}
Finally, let $\{\mut\}_{t\in \errep}$ be
the continuous convolution semigroup of measures associated with
the representation kit $\{\bj,\ajk,\eta \}_{j,k=1}^n$.
From formula~{(\ref{compa})} it follows that $\gene=\genera$.

The proof is complete.~{$\square$}

\begin{remark} {\rm
Given any pair of representation kits $\{\bj,\ajk, \eta
\}_{j,k=1}^n$ and $\{\tbj,\tajk, \tilde{\eta} \}_{j,k=1}^n$ (of
convolution semigroups of measures on $G$), for all
$r,\tilde{r}\in\errep$ one can define the set
\begin{equation}
r\cinque \{\bj,\ajk, \eta \}_{j,k=1}^n + \tilde{r}\cinque
\{\tbj,\tajk, \tilde{\eta} \}_{j,k=1}^n := \{r\quattro\bj +
\tilde{r}\quattro\tbj,r\quattro\ajk + \tilde{r}\quattro\tajk,
r\quattro\eta + \tilde{r}\quattro\tilde{\eta}\}_{j,k=1}^n,
\end{equation}
which is again the representation kit of a convolution semigroup of
measures on $G$. Then, from Theorem~{\ref{main-th}} it follows that
the set
\begin{equation}
\generau :=\big\{\genera\in\supops\colon\ \mbox{$\{\mut\}_{t\in
\errep}$ continuous conv.\ sem.\ of measures on $G$} \big\}
\end{equation}
of all generators of twirling semigroups associated with the
representation $U$ is a convex cone contained in $\coco(\suni)$.
Note that the convex cone $\generau$ is not `pointed' (i.e., it is a
`wedge'), unless the representation $U$ is trivial. In fact, we have
that
\begin{equation} \label{lineality}
\lineal :=\generau \cap (-\generau) =\big\{\ima
\big[\oph,(\cdot)\big]\colon\ \oph\in\vu\big\}.
\end{equation}
The set $\lineal$ is the `lineality space'~{\cite{Stoer}} of the
convex cone $\generau$. It is a vector space contained in the closed
cone $\zercone(\cvvv -I)$. The lineality space $\lineal$ is the
smallest face (extreme subset) of the convex cone $\generau$;
namely, it is a face of $\generau$, and any other face of $\generau$
contains $\lineal$. Moreover, the following decomposition holds:
\begin{equation}
\generau = \lineal + \generauno,
\end{equation}
where $\generauno$ is the pointed cone defined by $\generauno :=
\{0\}\cup (\generau\smallsetminus\lineal)$.~{$\blacksquare$} }
\end{remark}

Recalling the second assertion of Proposition~{\ref{twi-ran}}, and
applying the last assertion of Theorem~{\ref{main-th}} to the
defining representation of the group $\sunig(\dime)$, we get the
following result.
\begin{corollary}
Let $\hh$ be a finite-dimensional Hilbert space. Then, every
twirling semigroup acting in $\bH$ is a random unitary semigroup
and, conversely, every random unitary semigroup acting in $\bH$
arises as a twirling semigroup.
\end{corollary}

%------------------------------------------------------------------------------
\section{Conclusions, final remarks and perspectives}
\label{conclusions}
%------------------------------------------------------------------------------

In the present contribution, we have studied the main properties of
a well defined class of semigroups of (super)operators acting in
Banach spaces of trace class operators. These semigroups of
superoperators --- that we have called \emph{twirling semigroups}
--- are associated in a natural way with the pairs of the type
$(U,\{\mut\}_{t\in\errep})$, where $U$ is a projective
representation of a l.c.s.c.\ group $G$ and $\{\mut\}_{t\in\errep}$
is a continuous convolution semigroup of measures on $G$. In
Sect.~{\ref{twirling}}, we have proved that the twirling semigroups
are \emph{quantum dynamical semigroups}. Hence, they describe the
dynamics of a class of \emph{open quantum systems}. In order to
provide a characterization of this class of dynamical semigroups, we
have studied their infinitesimal generators.

As a first step, we have analyzed in detail the case where $G$ is a
Lie group and $U$ is a finite-dimensional, smooth (equivalently,
continuous), unitary representation. However, we stress that, thanks
to Nelson's theory of analytic vectors~{\cite{Nelson-anv}}, one can
extend some of the results of Sect.~{\ref{groups}} to the case where
$U$ is a generic strongly continuous unitary representation by
taking care of the domains of the (in general, unbounded)
infinitesimal generators of the associated twirling semigroups. This
task will be accomplished elsewhere~{\cite{Aniello-future}}.

The main technical tool that we have exploited for proving the main
result of Sect.~{\ref{groups}} --- i.e., Theorem~{\ref{main-th}} ---
is the classical L\'evy-Kintchine formula; but, as the reader will
have noticed, it has been necessary to prove Lemma~{\ref{biglem}} in
order to use this formula `as if the (smooth) function $G\ni
g\mapsto\langle \phi, U(g)\tre \opa\sei U(g)^\ast \psi\rangle$,
$\opa\in\bH$, $\phi,\psi\in\hh$, belonged to $\cdc(G;\ccc)$' (which,
of course, in general is not the case, unless $G$ itself is
compact). Moreover, as the reader may verify, to derive the
expression of the infinitesimal generator of the twirling semigroup
associated with the pair $(U,\{\mut\}_{t\in\errep})$ is simpler if
one assumes that $\{\mut\}_{t\in\errep}$ is a \emph{Gaussian}
semigroup of measures (to this aim, one can exploit the defining
condition~{(\ref{hypo})}); i.e., if $\{\mut\}_{t\in\errep}$ is the
distribution associated with a Brownian motion on $G$.

In addition to these technical remarks, it is also worth observing
that twirling semigroups are a natural source of \emph{covariant}
quantum dynamical semigroups. In fact, let $\{\twis\}_{t\in\errep}$
be the twirling semigroup associated with the pair
$(U,\{\mut\}_{t\in\errep})$. Consider the set
\begin{equation}
\gumt := \big\{g\in G\colon\
\twis\big(\urep(g)\opa\big)=\urep(g)\big(\twis\tre\opa\big),\
\forall\cinque t\in\errep,\ \forall\cinque\opa\in\trc\big\}.
\end{equation}
As the reader may easily check, $\gumt$ is a closed subgroup of $G$.
This subgroup includes the set
\begin{equation}
\gu := \{g\in G\colon\ \urep(gh)=\urep(hg),\ \forall\cinque h\in
G\},
\end{equation}
which is a closed normal subgroup of $G$ containing the center of
$G$. For instance, in the case where $U$ is a projective
representation of an \emph{abelian} group $G$, we have:
\begin{equation}
G=\gumt=\gu.
\end{equation}
Now, let $\subg$ be any subgroup of $\gumt$, and let
$\subre\colon\subg\rightarrow\uni(\hh)$ be the projective
representation defined by
\begin{equation}
\subre (g) = U(g),\ \ \ \forall\cinque g\in\subg.
\end{equation}
Then, we have that
\begin{equation}
\twis\big(\subre(g)\tre\opa\sette\subre(g)^\ast\big)
=\subre(g)\big(\twis\tre\opa\big)\tre\subre(g)^\ast,\ \ \
\forall\cinque t\in\errep,\ \forall\cinque g\in\subg,\
\forall\cinque\opa\in\trc;
\end{equation}
namely --- by definition, see~{\cite{Holevo}} --- the quantum dynamical semigroup
$\{\twis\}_{t\in\errep}$ is covariant
with respect to the representation $\subre$.

Another issue that is worth discussing is the characterization of
the twirling superoperators that are \emph{Markovian
channels}~{\cite{Wolf,Cirac}} (we would prefer the term
\emph{embeddable channels}); i.e.\ that are members of quantum
dynamical semigroups. Precisely, a twirling superoperator
$\mathfrak{S}$ is a Markovian channel if
$\mathfrak{S}=\mathfrak{S}_1$, for some quantum dynamical semigroup
$\{\twis\}_{t\in\errep}$ (not necessarily a twirling semigroup).
Clearly, if the twirling superoperator $\mathfrak{S}$ is associated
with a pair $(U,\mu)$ (which is, in general, not unique) such that
the probability measure $\mu$ is \emph{embeddable} --- namely,
$\mu=\mu_1$, for some continuous convolution semigroup of measures
$\{\mut\}_{t\in\errep}$ (see~{\cite{Heyer}}) --- then it is a
Markovian channel and a member of the twirling semigroup associated
with the pair $(U,\{\mut\}_{t\in\errep})$. However, whether
\emph{every} twirling superoperator which is a Markovian channel is
a member of a twirling semigroup seems to be an interesting open
problem. The investigation of this problem, in the light of known
results about the relation between  embeddable and \emph{divisible}
probability measures~{\cite{Heyer}}, may lead to a deeper
understanding of the relation between Markovian and \emph{divisible}
channels~{\cite{Cirac}}.

Finally, we note that, if the representation $U\colon
G\rightarrow\uni(\hh)$ is genuinely projective, by considering a
central extension~{\cite{Raja}} $\Gext$ of the circle group
$\mathbb{T}$ by $G$ one can always represent any twirling semigroup
associated with $U$ as a twirling semigroup associated with a
standard unitary representation of $\Gext$ (consider that every
convolution semigroup of measures on $G$ can be trivially extended
to $\Gext$).

%%%%%%--------------------------------------------------------------------------

%-----------------------------------------------------------------------------
\end{document}